\documentclass[a4paper,11pt]{article}
\usepackage[a4paper,
  left=2.5cm, right=2.5cm,
  top= 3cm, bottom=4cm]{geometry}
\usepackage{amsmath}
\usepackage{oldgerm}
\usepackage{amssymb}
\usepackage{bbm}
\usepackage[dvips]{graphicx}
\usepackage{epsfig}
\usepackage{color}
\usepackage{cite}
\usepackage{epic}

\newcommand{\rem}[1]{} 
\def\op{O}

\def\deg{\text{deg }}
\def\C{\mathbb{C}}
\def\Z{\mathbb{Z}}
\def\R{\mathbb{R}}

\def\aut{\operatorname{aut}}
\def\Aut{\operatorname{Aut}}
\def\nbo{N_{O\setminus B}}
\def\Hirz[#1]{\mathbbm{F}_{#1}}
\def\o[#1]{\overline{#1}}

\begin{document}

\thispagestyle{empty}
\vspace*{.2cm}
\noindent
HD-THEP-09-27 \hfill 7 December 2009 \\
ZMP-HH/09-32
\\

\vspace*{2.0cm}

\begin{center}
{\Large\bf D7-Brane Moduli vs. F-Theory Cycles\\ \vspace{.3cm} in Elliptically
Fibred Threefolds}
\\[1.5cm]
{\large A.~P.~Braun$^a$, S.~Gerigk$^b$, A.~Hebecker$^a$ and
H.~Triendl$^c$}\\[.5cm]
{\it$^a$ Institut f\"ur Theoretische Physik, Universit\"at Heidelberg,
Philosophenweg 16 und 19\\ D-69120 Heidelberg, Germany}
\\[.5cm]
{\it$^b$ Institut f\"ur Theoretische Physik, ETH Z\"urich, Wolfgang-Pauli-Strasse 27\\ CH-8093 Z\"urich, Switzerland}
\\[.5cm]
{\it$^c$ II. Institut f\"ur Theoretische Physik der Universit\"at Hamburg,
Luruper Chaussee 149\\ D-22761 Hamburg, Germany}
\\[.5cm]
{\small\tt (\,a.braun@thphys.uni-heidelberg.de}{\small ,} {\small\tt
\,gerigk@phys.ethz.ch}{\small ,} {\small\tt
a.hebecker@thphys.uni-heidelberg.de} {\small and} {\small\tt
\,hagen.triendl@desy.de)}
\\[2.0cm]
{\bf Abstract}\\
\end{center}

We study the space of geometric and open string moduli of type IIB compactifications from the perspective of complex structure deformations of F-theory. In order to find a correspondence, we work in the weak coupling limit and for simplicity focus on compactifications to 6 dimensions. Starting from the topology of D7-branes and O7-planes, we construct the 3-cycles of the F-theory threefold. We achieve complete agreement between the degrees of freedom of the Weierstrass model and the complex structure deformations of the elliptic Calabi-Yau. All relevant quantities are expressed in terms of the topology of the base space, allowing us to formulate our results for general base spaces.

\newpage

\tableofcontents

\section{Introduction}

Type IIB flux compactifications on Calabi-Yau orientifolds with D3 and D7-branes are
promising candidates for embedding the standard model of particle physics in string theory \cite{Blumenhagen:2006ci}.
At the same time, they offer mechanisms for inflation, supersymmetry breaking and fine-tuning of
the cosmological constant. The moduli of such compactifications are typically stabilized by a
combination of fluxes and non-perturbative effects \cite{Kachru:2003aw,
Balasubramanian:2005zx, Lust:2005dy, Lust:2006zg, Lust:2005bd, Denef:2005mm}.
Such models can also be described by the weak coupling limit of F-theory
compactifications \cite{Vafa:1996xn, Sen:1997bp}. The latter have more recently been used to
construct attractive GUT models, especially in situations where the weak
coupling limit can not be taken \cite{Beasley:2008dc,Beasley:2008kw,Donagi:2008ca, Donagi:2008kj,Blumenhagen:2008zz,Marsano:2009ym,Blumenhagen:2009up,Marsano:2009gv,Blumenhagen:2009yv}.

From the F-theory perspective, the deformations of both D7-branes and the type IIB orientifold are part
of the geometric moduli space of an elliptically fibred Calabi-Yau.
To study fluxes in F-theory models, one can use the duality to M-theory in which
four-form fluxes can be turned on. These four-form fluxes correspond to both brane and bulk
fluxes in the dual type IIB model. Although this provides a nice way to see
that brane moduli can be stabilized by fluxes, it is hard to
map families of elliptic Calabi-Yaus to the corresponding brane
configurations. For recent work on the D-brane superpotential and the map between complex structure moduli of
elliptic Calabi-Yau manifolds and the moduli of branes in F-theory, see e.g. 
\cite{Jockers:2008pe,Grimm:2008dq,Alim:2009rf,Jockers:2009mn,Alim:2009bx,Grimm:2009ef,Aganagic:2009jq,Grimm:2009sy,Jockers:2009ti}
(for an alternative approach see \cite{Walcher:2006rs,Morrison:2007bm,Walcher:2009uj,Li:2009dz}). This map has been worked out in detail 
for the simplest elliptic Calabi-Yau manifold, $K3$, in \cite{Braun:2008ua}.

In this work we present a hands-on approach to the parameterization of the
brane moduli space in the case of type IIB orientifold models compactified on complex surfaces.
In particular, we find a parameterization in terms of the periods of the elliptically fibred
Calabi-Yaus in the dual F-theory picture.

Type IIB orientifolds with two compact complex dimensions arise from involutions acting on $K3$. Involutions of $K3$ have been classified by Nikulin~\cite{Nikulin:1979,Nikulin:1983,Nikulin:1986}. The resulting base spaces $B$ are Fano surfaces or (blow-ups of) Hirzebruch surfaces.
The fixed point locus of the involution defines the O-plane.
The corresponding F-theory model is defined on the Calabi-Yau threefold that is constructed as an elliptic fibration over the base space $B$. 
The monodromy points of the fibration give the location of the branes in $B$ and thus the motion of branes corresponds to complex structure deformations of the fibration, i.e.\ complex structure deformations of the threefold. Note that the branes are located on holomorphic hypersurfaces and therefore their positions are described by holomorphic polynomials, i.e.\ the branes are given by divisors.
F-theory models on Calabi-Yau threefolds have been first discussed in~\cite{Morrison:1996na,Morrison:1996pp}.

By performing the weak coupling limit we can make contact with the corresponding orientifold model~\cite{Sen:1996vd,Sen:1997bp}.
The monodromy of D-branes and O-planes acts on the 1-cycles in the fibre torus. If one combines such a 1-cycle and an appropriate real surface with
boundary on the brane (a relative-homology cycle), one can construct a non-trivial 3-cycle. The deformation of such 3-cycles 
characterizes the deformation of the corresponding branes.

We are able to geometrically construct all 3-cycles in the way described above. We start at the orientifold point in moduli space, where the O-plane coincides with four D-branes, and construct the 3-cycles of the threefold describing the motion of the O-plane in the base. Then we move one D-brane after the other off the O-plane and construct the emerging cycles corresponding to their motion. By counting the degrees of freedom we see that we find all 3-cycles. 

We now give an overview of this work including the main results of each
section.

In Sect.~\ref{localConstruction} we start by investigating the recombination
of branes in a small neighborhood around an intersection point.
Geometrically, the recombination process blows up the nodal
point\footnote{At a nodal point an embedded Riemannian
surface is locally described by the equation $xy=0$ with $x,y \in \C$. This
gives rise to two intersecting hypersurfaces situated at $x=0$ and $y=0$.} at the
intersection, which generates a 1-cycle of the recombined brane with
non-vanishing size. This 1-cycle is the boundary of a disc: a relative 2-cycle. 
By fibering this disc with the 1-cycle of $T^2$ that degenerates on the boundary, 
these 1-cycles are shown to be in one-to-one correspondence to F-theory 3-cycles 
in the case of D-branes. In the case of O-planes, each such 1-cycle corresponds to 
two 3-cycles of the underlying F-theory threefold. The periods associated to these 
F-theory cycles determine the recombination of D-branes and O-planes.

We start addressing global issues in Sect.~\ref{DonO}. For simplicity, we
first restrict ourselves to the subset of elliptically fibred Calabi-Yau
spaces $Z$ which correspond to type IIB models at the orientifold point.
This means that each O-plane coincides with a stack of four D-branes. From
the geometric point of view, the brane locus is described by a Riemannian
surface $\cal D$ embedded in the base space $B$. According to the analysis of
Sect.~\ref{localConstruction}, the 1-cycles of $\cal D$ correspond to
3-cycles of $Z$ which parameterise the motion of the O-plane. In particular, a
self-intersection point of this O-plane develops if any of these 3-cycles
shrinks. We will refer to these cycles as 'recombination cycles'. We find that,
in addition to recombination cycles, there exist F-theory 3-cycles which are
associated to the 2-cycles of the base space $B$ of the type IIB model.
Locally these cycles can be visualized as products of the 2-cycles in
the base and a 1-cycle of the $T^2$ fibre. Combining them with the
recombination cycles, we have enough periods to parameterise the complex
structure of the type IIB orientifold model. Thus we have constructed all
3-cycles of $Z$ which have non-zero volume at the orientifold point.

A first step towards the generic situation is taken in Sect.~\ref{D7-branes without obstructions}.
First, we separate only one of the D-branes from the D-brane stack, leaving three
D-branes on top of the O-plane. Geometrically, the situation is appropriately
described by two hypersurfaces of the same degree embedded in $B$, which
generically intersect each other in isolated points. Fixing the O-plane in $B$,
we then identify loci in the base which correspond to F-theory
3-cycles. In contrast to the cycles governing the deformations of the O-plane, we
end up with cycles of two different kinds. The first kind of cycle is a relative
2-cycle stretched between a 1-cycle of the D-brane and a 1-cycle of the O-plane.
It locally measures the distance between D-brane and O-plane.
The second kind of cycle is again a relative 2-cycle. By contrast to the first kind, 
its boundary is not formed by 1-cycles in D-brane and O-plane but by two lines 
connecting a pair of O-plane-D-brane intersection points. One may think of
this 2-cycle as measuring both the distance between D-brane and O-plane 
and the distance between two of their intersection points.

Next, we consider more general D-brane configurations, demanding only that at least 
one D-brane remains on top of the O-plane. In this case, D-brane deformations are still 
associated to deformations of generic hypersurfaces.
To be more explicit, the D-brane locus $\eta^2+h\chi=0$, in the
notation of \cite{Sen:1997bp}, can be restricted to be of the form
$\eta=hp$, which yields $h(hp^2+\chi)=0$. This
corresponds to a situation, in which one D-brane coincides with the
O-plane, but all other D-branes are recombined into the generic surface $\chi'=hp^2+\chi=0$.
In this case we can construct a complete base of $H_3(Z)$ by iteratively moving single 
D-branes independently off the O-plane and letting them recombine at their intersection 
points according to the results in Sect.~\ref{localConstruction}.

In Sect.~\ref{moreDoffO} we finally discuss the most general case of a `naked' O-plane and a fully
recombined single D-brane. The latter can only have double intersections 
with the O-plane and is hence no longer given by a generic hypersurface \cite{Braun:2008ua, Collinucci:2008pf}. 
We compute the number of moduli in three ways. First we count the number of deformations that are contained in a 
polynomial of the form $\eta^2+h\chi=0$ for any given base space. We show that the difference between this
number and the number of moduli of a generic hypersurface is given by the number of double-intersection points.
We also compute the number of moduli by describing the deformations as
sections of a particular line bundle over the D-brane (which is a $\mathbb{Z}_2$-twist of the canonical bundle).
Finally, we show that our construction yields precisely the right number of cycles to explain the
degrees of freedom from the perspective of the elliptic threefold.

We end this work with conclusions and an outlook on possible applications and
generalization to compactifications to four dimensions. The appendix contains
some technical details of the methods used in this work.

\section{Local construction} \label{localConstruction}

We are interested in the recombination and displacement of O7-planes
and D7-branes in complex two-dimensional type IIB orientifolds. In the picture
given by F-theory, these moduli are encoded in complex structure deformations of
the corresponding elliptically fibred Calabi-Yau threefold. As these complex
structure deformations originate from 3-cycles we are interested in finding
these. To begin our analysis, we start constructing the threefold cycles in the
weak coupling limit locally from the topology of the D-branes and O-planes and
our knowledge of the fibration. In a similar fashion as in \cite{Braun:2008ua},
we will describe these cycles as a fibration of a 1-cycle in the fibre over some
(real) surface in the base. Instead of considering the whole threefold, we will
first consider O7-plane/D7-brane configurations in flat space. As we know the
elliptic fibration over these configurations, this will give a local picture of
the elliptically Calabi-Yau threefold in which we can identify some of the
3-cycles.

\subsection{Recombination of two intersecting D7-branes} \label{recomD}

\begin{figure}
\begin{center}
\includegraphics[height=4cm, angle=0]{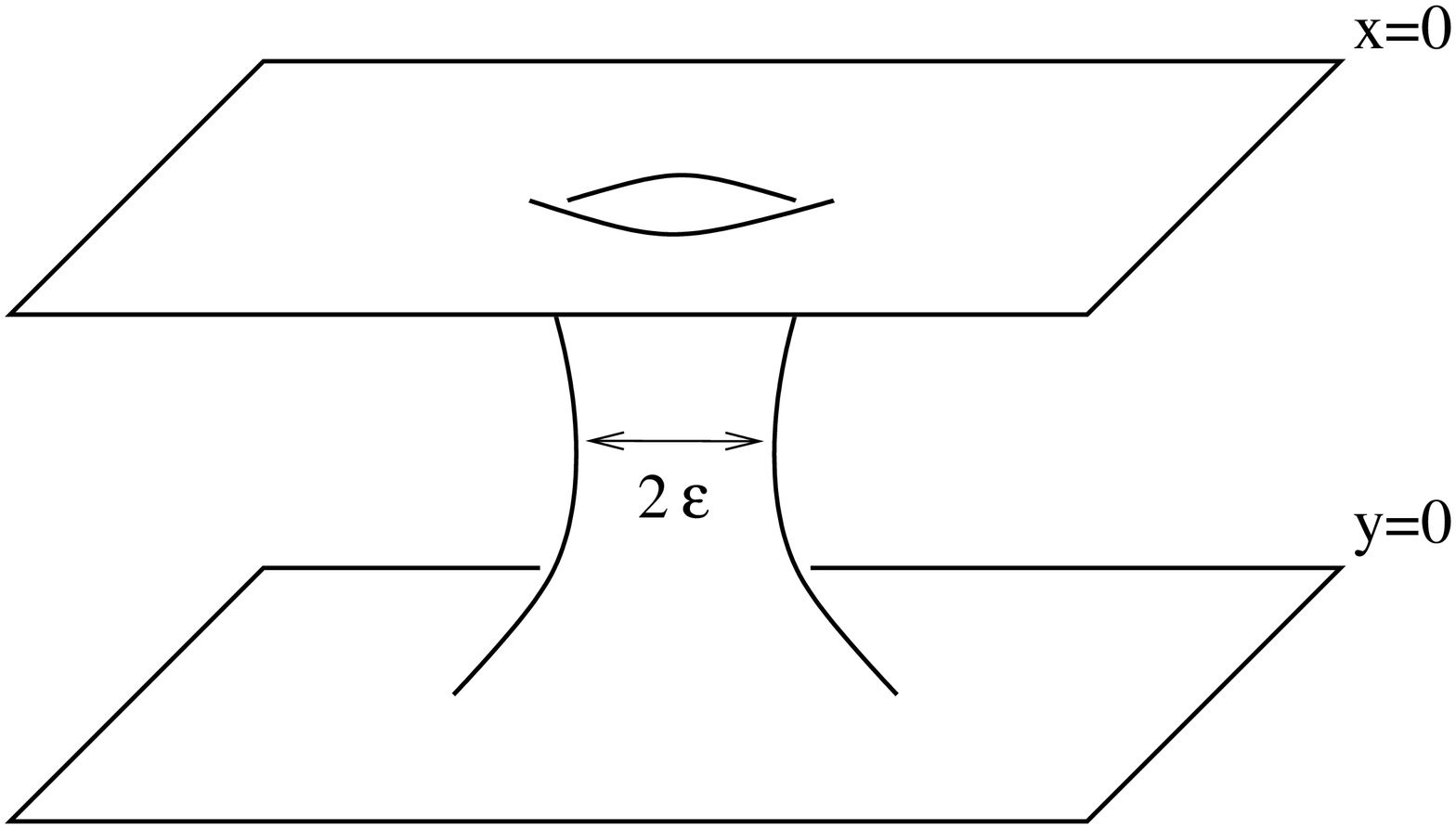}
\end{center}
\caption{\textsl{The surface formed by the recombined D-brane, as described by
eq.~(\ref{2D}). The parameter $\epsilon$ determines the radius of the circle
that sits at the narrowest point.}}
\label{hyperbolo}
\end{figure}

Consider two D-branes in a complex $2$-dimensional base space the intersection point of which 
is well-separated from other branes and O-planes. This situation is described by the equation
\begin{equation}
xy=0 \ ,
\end{equation}
which factorizes into $x=0$ and $y=0$.
The recombination is characterized by the deformation
\begin{equation}
xy=\tfrac{1}{2} \epsilon^2  \ ,  \label{2D}
\end{equation}
after which the equation no longer factorizes.
Far away from the intersection, for $|x|\gg\epsilon$ or
$|y|\gg\epsilon$, the recombined brane is still approximated by two branes at $y=0$ and
$x=0$. We take $\epsilon$ to be real. To understand the topology of the recombined D-brane, we introduce new coordinates
\begin{equation}
x=r\exp{\mbox{i}\phi} \ , \hspace{1cm} y=\rho\exp{\mbox{i}\psi} \ .
\end{equation}
In these coordinates, eq.\eqref{2D} reads
\begin{equation}
r\rho=\tfrac{1}{2} \epsilon^2 \ , \hspace{1cm} \phi+\psi=0 \ . \label{2D_polar}
\end{equation}
The equation
\begin{equation}
r=r_0
\end{equation}
characterizes an $S^1$ parameterized by $\phi \in [0,2\pi)$. When $r_0$ varies between zero and infinity, this loop sweeps out the whole recombined D-brane. The length of this loop, which is given by $2\pi\sqrt{r_0^2+\left(\frac{\epsilon^2}{2r_0}\right)^2}$, diverges when $r_0$ tends to zero or infinity, corresponding approximately to a circle in a brane at $x=0$ or $y=0$. It takes its minimum value, $2\pi \epsilon$, for $r_0= \tfrac{1}{\sqrt{2}} \epsilon$. The
topology of the recombined brane is thus given by a `throat' that connects two
asymptotically flat regions, as shown in Figure \ref{hyperbolo}.

\begin{figure}
\begin{center}
\includegraphics[height=4cm, angle=0]{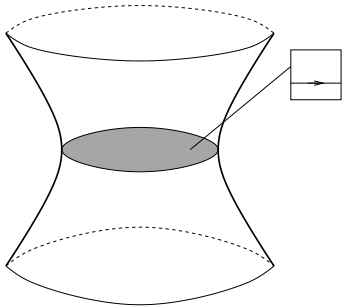}
\end{center}
\caption{\textsl{By taking a disc which has its boundary on a D-brane and
adding the horizontal cycle in the fibre torus at every point, we obtain a non-trivial 3-cycle.}}
\label{3S}
\end{figure}

When $\epsilon\rightarrow 0$, the `minimal loop' $r_0=\tfrac{1}{\sqrt{2}} \epsilon$ collapses and
eq.~\eqref{2D} factorizes, corresponding to two intersecting D-branes. We can
now construct the $3$-cycle that controls this process from the F-theory point of
view:
We recall that, over every point of the four-dimensional base space in which the brane is embedded, we have a torus fibre and that the $(1,0)$-cycle of this torus shrinks at the D7-brane locus. Consider a disc in the base space the boundary of which is the $1$-cycle of the D7-brane world volume discussed above (e.g.\ with $r_0=\tfrac{1}{\sqrt{2}} \epsilon$). The relevant $3$-cycle is obtained by taking the $(1,0)$-cycle of the fibre torus at every point of this disc.
This is illustrated in Figure~\ref{3S}. One easily convinces oneself that this cycle is a $3$-sphere.
It is obvious from the above that the volume of this $3$-cycle, divided by the square root of the fibre 
volume to keep it finite in the F-theory limit, characterizes the 
recombination process.\footnote{The F-theory limit of M-theory is characterized by the limit of zero size of the elliptic fibre. Since the complex structure moduli of the threefold should be independent of the size of this 2-cycle, we have to rescale the volume of 3-cycles by appropriate powers of its size.}

\subsection{Recombination of two intersecting O7-planes}\label{recO}

In the following we will locally construct 3-cycles in F-theory that
correspond to the movement of O-planes in its Type IIB dual. In order to simplify
the monodromy structure, we will consider the (singular) case of four D-branes coinciding
with the O-plane. In this way, the only monodromy appearing is an involution of
the fibre torus.

The recombination of two O7-planes is described by the same equation~\eqref{2D} as in
the D7-brane case. Thus two recombined O7-planes will
also form a surface which contains a throat supporting a circle of minimal
circumference. 

\begin{figure}[h]
\begin{center}
\includegraphics[width=8cm]{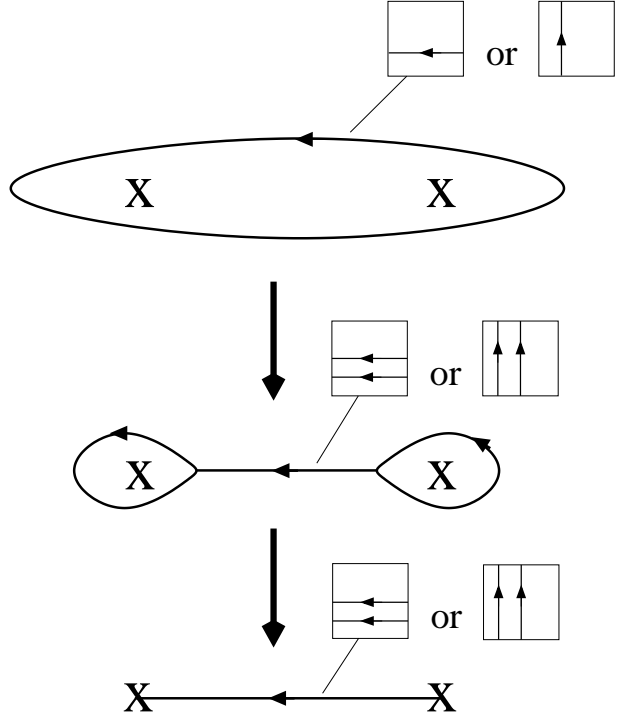}
\end{center}
\caption{\textsl{O-planes in a complex two-dimensional base give rise to cycles that have an arbitrary component in
the fibre and encircle the positions of the two O-planes. As shown in the figure, one can subsequently deform these 
cycles so that their base component becomes a line connecting the two O-planes. As the fibre component changes its
orientation upon circling one of the O-planes, the fibre component of the resulting line is twice that of the
original loop.}}
\label{o7cyc}
\end{figure}

To describe the cycle that controls the recombination of the O7-plane, let
us first recall its construction in the case of a complex one-dimensional base
space. In this case, the O-planes are merely points in the base. We can construct
a non-trivial cycle by taking a loop that circles two O-planes, together with an
arbitrary component in the fibre. As is shown in Figure \ref{o7cyc}, we can collapse
this cycle to a line that starts at one of the O-planes and ends at the other one.

Keeping the construction in the case of a complex one-dimensional base in mind, we can repeat the construction done 
for the D-brane: we take a disc ending on the O-plane in the base and one of the two fibres in Figure \ref{o7cyc}
to construct a 3-cycle. Just as in the D-brane case, the size of this cycle will describe the recombination
process of two intersecting O-planes.

\section{F-Theory models at the orientifold point} \label{DonO}

\begin{figure}
\begin{center}
\input{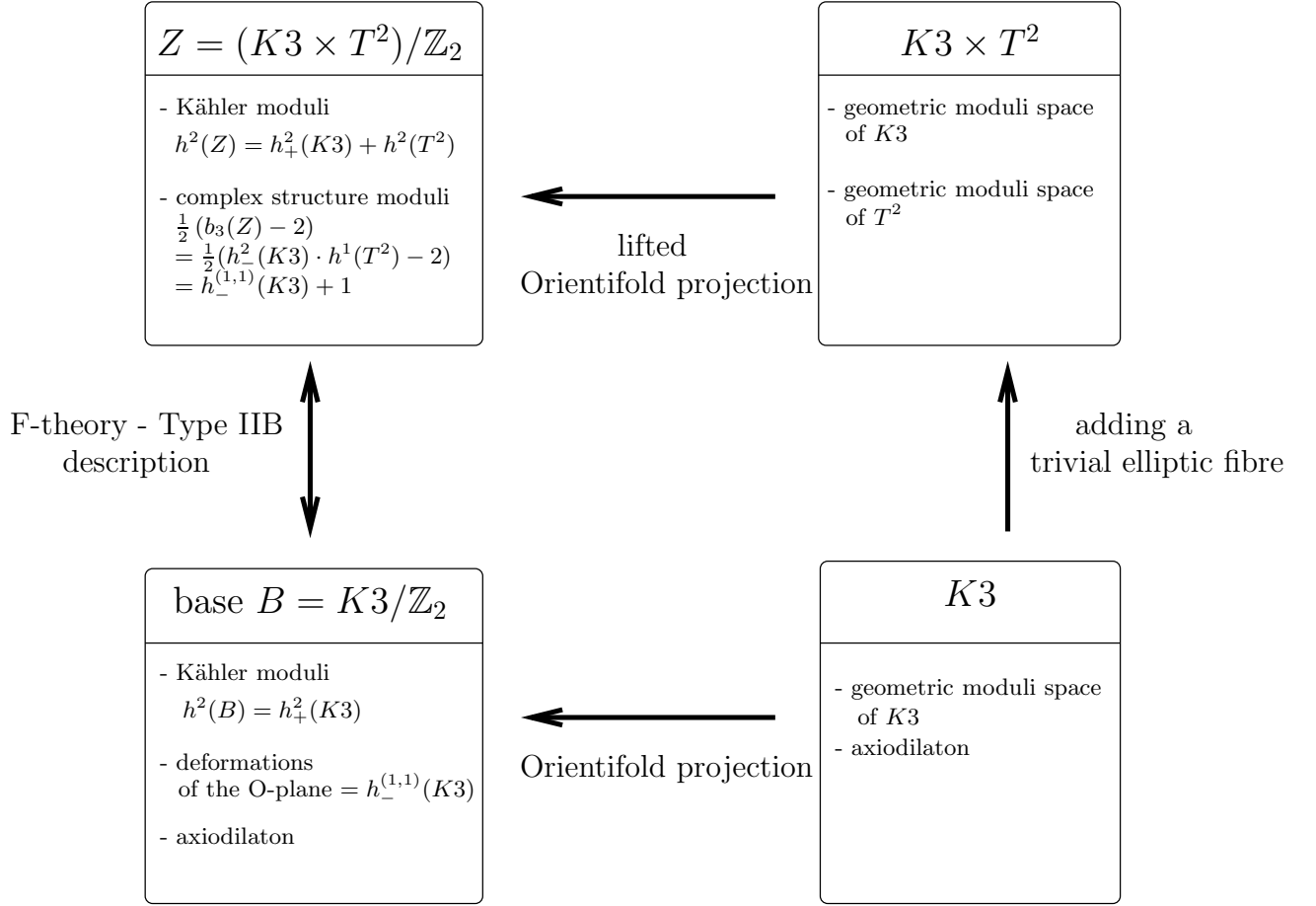}
\end{center}
\caption{\textsl{Type IIB orientifolds of $K3$ may be described by F-theory on
$Z=(K3\times T^2)/\Z_2$. The complex structure deformations of $Z$ correspond to
deformations of the O-plane and the axiodilaton.}} \label{scheme}
\end{figure}

In the following, we discuss F-theory compactifications on elliptic threefolds that can be
constructed as orientifolds. In particular, we demand that all D-branes coincide with the O-plane in this
section. We are going to describe these models from several perspectives, summarized
 in Figure \ref{scheme}.

\subsection{The type IIB perspective}\label{orientiIIB}

Let us start with the well-known type IIB perspective. Besides the various form-fields, the moduli
space of type IIB on $K3$ contains the geometrical moduli space of $K3$ and the complexified
string coupling, also known as the axiodilaton (see e.g. \cite{Aspinwall:1996mn}). The geometric
moduli can be elegantly described as the rotations of a three-plane of positive norm inside
$H^2(K3)$. The three positive-norm vectors that span this three-plane can then be used to construct
the K\"ahler form $J$ and the holomorphic two-form $\Omega^{(2,0)}$. As is common for Calabi-Yau threefolds,
the geometric moduli of $K3$ can then be mapped to K\"ahler and complex structure deformations.

Apart from the inner parity of the various degrees of freedom coming from the involution of the
world-sheet, orientifolding includes an involution $\iota$ of space-time. Furthermore, this involution
has to map the holomorphic two-form $\Omega^{(2,0)}$ to minus itself, so that the quotient space
$B$ is not Calabi-Yau. Involutions of this kind are known as \emph{non-symplectic} in the mathematics
literature and have been classified by Nikulin \cite{Nikulin:1986}. We have summarized the main
results in Appendix \ref{nikulinClassification}. The classification of non-symplectic involutions of
$K3$ implies that $B$ is rational, so that it is a del Pezzo surface $dP_i$, a Hirzebruch surface $\Hirz[n]$,
or a blow-up of a Hirzebruch surface. For a short review of rational surfaces see Appendix~\ref{complexSurfaces}.
The mapping between type IIB orientifolds and the base space of the corresponding F-theory model has
recently been discussed in detail in \cite{Collinucci:2008zs,Collinucci:2009uh}.

Under the action of $\iota$, the cohomology groups of $K3$ decompose into eigenspaces:
\begin{equation}
H^{(p,q)}(K3)=H^{(p,q)}_+(K3)\oplus H^{(p,q)}_-(K3)\ .
\end{equation}
The geometric moduli of this orientifold model were discussed in detail in
\cite{Brunner:2003zm}. The complex structure deformations that are compatible
with $\iota$ are in one-to-one correspondence with elements of $H^{(1,1)}_-(X)$.
As the K\"ahler form of $K3$ is even under $\iota$, compatible K\"ahler deformations
can be parameterized by $H^{(1,1)}_+(K3)$. Since $\iota$ is a non-symplectic involution, we have
$H^{(1,1)}_+(K3)=H^2_+(K3) \simeq H_2^+(K3)=H^2(B)$.

The fixed point locus of $\iota$ is the orientifold plane $O$.
It is given by the vanishing locus of a section of $[-2K_B]$, where $[K_B]$ denotes the
canonical bundle\footnote{For a divisor $D$ we denote the corresponding line bundle by $[D]$.}
of the base space \cite{Sen:1996vd}. In other words, the O-plane is equivalent to $-2K_B$ as a
divisor. This is necessary to ensure that the double cover is a Calabi-Yau space. To cancel the D7-brane
charge, the homology class of all the D-branes has to equal four times the
homology class of the O-plane. In this section we choose to align four D-branes with the O-plane. The only
geometric deformations of this configuration are hence given by the K\"ahler deformations of the base and
the deformations of the O-plane. These must be equivalent to the deformations
of $K3$ compatible with the orientifolding.

Let us illustrate this in the simple example of $B=\C P^2$. The sections of $[-2K_{\C P^2}]$ are
given by homogeneous polynomials of degree 6. We can count the degrees of freedom that correspond to
deformations of this polynomial: there are 28 independent monomials and hence 28 complex coefficients.
One of these can be set to unity by an overall rescaling. In addition, the embedding of $O$ in
$\C P^2$ is only defined up to automorphisms of $\C P^2$. This automorphism group is complex 8-dimensional
(see Appendix \ref{toricAut}), eliminating $8$ degrees of freedom. We thus end up with $19$ complex degrees of
freedom. As $b_2(\C P^2)=1$, we find that $h^{(1,1)}_-=19$, giving the right number of degrees of freedom.
We have collected some more examples in Table \ref{summExam} at the end of the present section.

\subsection{Deformations of the O-plane}

Deformations of a Riemannian surface, such as the O-plane $O$, are
given by holomorphic sections of its normal bundle $[\nbo]$. The dimension of the
space of holomorphic sections of $[\nbo]$ is commonly denoted by $h^0(\nbo)$.

A Riemannian surface, such as the O-plane, has $3g(O)-3$ complex structure deformations. Let us explain why this is also the number of deformations in the embedding.
From the adjunction formula we have $K_O=\nbo+K_B$.\footnote{Here and in the following the restriction of $K_B$ to $O$ is implicit.}
As $O$ is linearly equivalent to $-2K_B$, we have $\nbo = -2K_B$ and thus find $\nbo = 2K_O$.
Serre's duality then tells us that
\begin{equation}
H^0(\nbo)= H^0(2K_O)= H^1(T_O)^* \ ,
\end{equation}
in other words we find
\begin{equation}
h^0(\nbo) = 3g(O)-3\ . \label{moduliRiemann}
\end{equation}

For later convenience, we show how to derive \eqref{moduliRiemann} using the Riemann-Roch-Theorem \cite{Griffiths:1994}
\begin{equation}
h^0(\nbo) = h^0(K_O-\nbo)+\text{deg } \nbo - g(O) + 1
\end{equation}
where $K_O$ is the canonical divisor of $O$ and $g(O)$ the genus of $O$.
The degree of a line bundle $L$ is the number of zeros of a generic section of
$L$. If $\text{deg } \nbo > \text{deg } K_O = 2g(O)-2$ it follows that $h^0(K_O-\nbo)
= 0$. In this case
\begin{equation}
h^0(\nbo) = \text{deg } \nbo - g(O) + 1. \label{dimHolSections}
\end{equation}
Since a section of the normal bundle $[\nbo]$ is nothing but a
deformation of the Riemannian surface, $\text{deg } \nbo$ is just the
self-intersection number of $O$.
As $O$ is linearly equivalent to $-2K_B$, we find that its self-intersections number is
$O\cdot O=4 K_B\cdot K_B$. Furthermore, we find from $K_O = -K_B$ the Euler characteristic of $O$
to be $\chi_O = \nbo \cdot K_O= -2K_B^2$.
Hence we obtain
\begin{equation}
\text{deg } \nbo = 4 K_B\cdot K_B = -2 \chi_O = 4g(O)-4 > 2g(O)-2 \text{   for }
g(O) \geq 2 \ .
\label{selfintersection}
\end{equation}
Thus the requirement for Eq.\ \eqref{dimHolSections} is satisfied and we find \eqref{moduliRiemann}.

So far, we have neglected the fact that some deformations of the O-plane are equivalent
to applying an automorphism $A \in \Aut(B)$ of the base $B$ and as such do not
represent valid complex degrees of freedom. Thus the number of deformations of the O-plane, $\text{Def }O$,
is given by
\begin{equation}
\dim_\C\text{Def}O = 3g-3 - \dim_\C \aut_B\ ,
\end{equation}
where $\aut_B$ denotes the Lie-algebra of $\Aut(B)$. In the
case of a toric variety, this quantity can be found by the procedure
explained in Appendix \ref{toricAut}.

\subsection{F-theory perspective}\label{fthpers}

Having discussed the moduli from the type IIB perspective, we now turn to the
F-theory description of the same situation. Since we have taken all D-branes
to be aligned with the O-plane, the axiodilaton is constant along $B$.
The corresponding F-theory description thus must be such that the complex structure of the
elliptic fibre is constant. Before orientifolding, there are no $SL(2,\Z)$ monodromies and
the F-theory threefold is simply the product $K3\times T^2$. The orientifolding introduces
a monodromy that acts as an involution on the fibre $T^2$. It occurs upon encircling the O-plane
locus. We can describe this situation by lifting the involution $\iota$ to an involution $\tilde{\iota}$
on $K3 \times T^2$ by defining
\begin{equation}
\tilde{\iota}(x,z)=(\iota x, -z)\ , \label{Finvolution}
\end{equation}
where $x \in X$ and and $z \in T^2$. Modding out $\tilde{\iota}$ yields the
F-theory compactification on $Z=(K3\times T^2)/\Z_2$.

The homology groups of $K3 \times T^2$ are
\begin{eqnarray*}
H_1(K3 \times T^2) &=& H_1(T^2) \ , \\
H_2(K3 \times T^2) &=& H_2(K3) \oplus H_2(T^2)\ , \\
H_3(K3 \times T^2) &=& H_2(K3) \otimes H_1(T^2)\ ,
\end{eqnarray*}
since $H_1(K3) = 0$.
Keeping the cycles even under $\tilde{\iota}$ yields the homology of $Z$:
\begin{eqnarray}
H_2(Z) &=& H_2^+(K3) \oplus H_2(T^2)\ , \nonumber \\
H_3(Z) &=& H_2^-(K3) \otimes H_1(T^2) \ . \label{homologyZ}
\end{eqnarray}

F-theory on $Z$ emerges from M-theory on the same manifold in the limit of
vanishing fibre size. The geometric moduli of M-theory on $Z$ are deformations of $X\times T^2$
that respect the $\Z_2$ action. Hence the K\"ahler and complex structure deformations of $Z$ are
linked to even cycles of $K3 \times T^2$. The K\"ahler moduli of $B$ and the volume of the elliptic
fibre become the K\"ahler moduli of $Z$. As the fibre size tends to zero in the F-theory limit, it does
not give rise to a physical modulus in F-theory, so that we find the same number of K\"ahler moduli as for the $K3$-orientifold $B$.

The 3-cycles of $Z$ originate from the odd 2-cycles of $K3$.
The number of complex structure moduli, $\dim_\C {\cal M}^{CS}$, of a Calabi-Yau space is given by the
number $h^{(2,1)}=\frac{1}{2}\left(b^3-2\right)$ \cite{Candelas:1990pi}. If we choose a
symplectic basis $(A^a, B_b)$, the complex structure moduli space is locally parameterized by the $h^{(2,1)}$
independent periods:
\begin{equation}
z^a=\int_{A^a}\Omega,\hspace{1cm} \Pi_b(z)=\int_{B_b} \Omega \ ,
\end{equation}
where $\Omega$ is the holomorphic 3-form. In the present case, we only
consider complex structure deformations that do not destroy the structure $Z=(K3\times T^2)/\Z_2$ (by resolving the orbifold singularities, for instance).
We can think of this restriction as fixing
a number of periods. The relation between complex structure deformations and $b^3$,
$\dim_\C {\cal M}^{CS}=\frac{1}{2}\left(b^3-2\right)$ thus holds for an orbifold like $Z$ as well.

In the present case we find that
\begin{equation}
\dim_\C {\cal M}_Z^{CS}=\frac{1}{2}\left(b^3(Z)-2\right)=\frac{1}{2}\left(2b^{2}_{-}(K3)-2\right)=b^{2}_-(K3) -1\ .
\end{equation}
As the holomorphic two-form of $K3$ and its complex conjugate are always
odd for the involutions considered we have $b^2_-(K3)=h^{(1,1)}_-(K3)+2$, so that we find
\begin{equation}
\dim_\C {\cal M}_Z^{CS}=h^{(1,1)}_-(K3) +1\ . \label{csZ}
\end{equation}

In F-theory, the present situation is described by an elliptic threefold in which
there is a $D_4$ singularity along a curve that is equivalent to $-2K_B$. This
curve is the location of the O-plane. The complex structure deformations of the
threefold that preserve the singularity structure correspond to deformations
of the O-plane and the value of the axiodilaton. This is expressed in~\eqref{csZ}:
the number of complex structure deformations equals the number of deformations of the
O-plane, $h^{(1,1)}_-(K3)$, plus one. This fits nicely with the aforementioned result that
deformations of the double cover $K3$ compatible with the orientifold involution originate
from cycles of $K3$ that are odd under the orientifold involution.

There is a subtle point worth mentioning here. In the present case, $Z$ has four $A_1$ singularities along the
location of the O-plane. The manifold constructed from the Weierstrass model that corresponds to the orientifold $B$ has however a $D_4$ singularity over the location of the O-plane. Even though for both the fibre torus
undergoes the same monodromies \cite{Friedman-Morgan(1994)}, they give rise to different physics
for compactifications of M-theory. However, in the F-theory limit both manifolds coincide, as
they are connected by blow-ups and blow-downs of the singular fibres, leading to the same type IIB model. See \cite{Braun:2009wh} for a recent discussion of this and related issues.

\subsection{3-cycles at the orientifold point}
In the following, we will discuss how the 3-cycles of $Z$ emerge from the topology
of the O-plane. Let us first return to the example of $B=\C P^2$. In this case
the O-plane locus has $19$ deformations, so that we expect the corresponding threefold
to have $20$ complex structure moduli yielding $42$ independent 3-cycles.

The local construction of F-theory 3-cycles from the O-plane topology, presented in Section \ref{recO},
suggests that we obtain two F-theory cycles
for each 1-cycle of the O-plane $O$. Since $\dim H_1(O)=2g(O)$, where $g(O)$ is
the genus of $O$, we expect $4g(O)$ F-theory 3-cycles. As mentioned before, $O$
is given by a defining polynomial $h$ of degree 6 in the $\C P^2$ case,
yielding a curve of genus $g(O)=10$. We thus obtain 40 cycles in $H_3(Z)$ instead of 42.
However, from the global point of view there is exactly one other cycle that could be lifted
to a F-theory 3-cycle, namely the cycle corresponding to the
hyperplane divisor $H$ of $\C P^2$. Naively, we can add a leg with two possible orientations
in the fibre to $H$ so that the lift yields two extra cycles. Including these, we get the
right number of 42 cycles. We have collected some more examples in Table \ref{summExam}.

There is one potential difficulty to this construction. The
hyperplane divisor will generically intersect the O-plane $O$. Since the
fibre degenerates on $O$ it is not a priori clear how to lift $H$ properly.
As it will become clear later on, $H$ can indeed be lifted to an F-theory cycle
but for now this remains a conjecture motivated by the counting.

It is now natural to conjecture that for any base space $B$, the number of non-degenerate
cycles in $H_3(Z)$ is given by
\begin{equation}
b_3(Z)=4g(O)+2 b_2(B)\ . \label{conjecture}
\end{equation}

The Lefschetz fixed point theorem  \cite{Griffiths:1994} allows us to relate the topology
of the O-plane to the topology of $Z$ in the general case. For a $K3$ surface it reads
\begin{equation}
2+b_2^+(X)-b_2^-(X)=\chi(O) \ . \label{Lefshetz2}
\end{equation}
From this is follows directly that
\begin{equation}
b_3(Z)=2b_2^-(K3)=4-2\chi(O)+2b_2^+(K3) \ .
\end{equation}
On the other hand, we know that the fixed point locus is given by the disjoint union of
a curve of genus $g$ and $k$ spheres \cite{Nikulin:1986}, see also Appendix \ref{nikulinClassification}.
Thus we have
\begin{equation}
\chi(O)=2-2g+2k \ ,
\end{equation}
which yields
\begin{equation}
b_3(Z)=4g+2b_2(B)-4k \ .     \label{relation}
\end{equation}
This equation can also be derived directly using \eqref{Nikulin}. For the cases in which the O-plane is given
by a single smooth complex surface we have $k=0$, so that (\ref{conjecture}) indeed holds. Note that this is
the case in the example of $B=\C P^2$ discussed before.

\begin{table}
\begin{center}
\begin{tabular}{|c||c|c|c|c|c|c|}
\hline
surface $B$ & $dP_0$ & $dP_1$ & $dP_2$ & $dP_3$ &
$\Hirz[0]$ & $\Hirz[2]$ \\
\hline
$g(O)$ & 10 & 9 & 8 & 7 & 9 & 9 \\
$b_2(B)$ & 1 & 2 & 3 & 4 & 2 & 2 \\
$h^{(1,1)_-}$ & 19 & 18 & 17 & 16 & 18 & 17 \\
$\dim_\C$ Def($O$) & 19 & 18 & 17 & 16 & 18 & 17 \\
\hline
$b_3(Z)=2h^{(1,1)}_-+4$ & 42 & 40 & 38 & 36 & 40 & 38 \\
\hline
$4g(O)+2b_2(B)-4k$ & 42 & 40 & 38 & 36 & 40 & 40 \\
\hline
\end{tabular}
\end{center}
\caption{\textsl{As discussed in the main text, the moduli of type IIB orientifold models on $K3$
can be described in different ways, see also Figure \ref{scheme}. This table contains some numerical
examples for simple base spaces $B$. The calculation of the appearing quantities is explained in the
text. Note that $dP_0= \C P^2$, $\Hirz[0] = \C P^1 \times \C P^1$ and $dP_1 =\Hirz[1]$.}}
\label{summExam}
\end{table}

The results for different surfaces are given in Table \ref{summExam}. It is interesting to note that in the case of
$\Hirz[2]$ the degrees of freedom in the type IIB picture do not fit the number
of complex structure moduli in the F-theory picture. Indeed, in this case
one can show that there is one complex structure deformation of the
Calabi-Yau threefold which is not realized as a polynomial deformation of the
Weierstrass model (see e.g. \cite{Green:1987rw} for a discussion of this
phenomenon in the physics literature).

It is nice to see that (\ref{relation}) is invariant under blow-ups of the base:
As
\begin{equation}
K_O=K_B+O=-K_B\ ,
\end{equation}
we deduce the Euler characteristic of $O$ to be
\begin{equation}
\chi(O) = -K_O\cdot O = -2 (-K_B)\cdot (-K_B)\ .
\end{equation}
Now consider the blow-up $\pi:\tilde{B} \rightarrow B$ at a generic point
$u\in B$. This means we add an exceptional divisor $E$ so that
$b^2(\tilde{B})=b^2(B)+1$. The behavior of the anticanonical divisor
under blow-ups is~\cite{Griffiths:1994}
\begin{equation}
-K_{\tilde{B}}=\pi^*(-K_B)- (\dim_\C B-1)E = -K_B-E \ .
\end{equation}
The exceptional divisor can always be chosen to satisfy \cite{Safarevic}
\begin{eqnarray}
E^2=-1 && E \cdot T_i = 0 \text{ for all toric divisors } T_i \in \text{Div}(B) \ .
\end{eqnarray}
In particular, this implies $-K_B \cdot E = 0$. It is now straight forward to
determine the Euler characteristic of $\tilde O$:
\begin{equation}
\chi(\tilde O)=-2\left(-K_B-E\right)\cdot\left(-K_B-E\right)=\chi(O)+2 \ .
\label{EulerBlowUp}
\end{equation}
Since the Euler characteristic is given by the relation $\chi=2-2g+2k$, eq. (\ref{EulerBlowUp})
implies that $g-k\rightarrow g-k-1$ under blow-ups of $B$ at generic
points. We thus need to show that $b_3(Z)\rightarrow b_3(Z)-2$ under blow-ups of the base.
As $b_2^-+b_2^+$ is fixed and the blow-up increases $b_2^+$ by one, $b_2^-$ must
decrease accordingly. Hence we find $b_3(Z)\rightarrow b_3(Z)-2$, so that eq.(\ref{relation})
remains valid.

Formula \eqref{relation} can be given a further interpretation in terms of the double cover $K3$. Let us start with the case $k=0$ and consider a 1-cycle of the O-plane. This 1-cycle is trivial inside the base. Thus, there exists a real disk in the base whose boundary coincides with this 1-cycle. If we now go to the double cover, we end up with two disks that are glued together at their boundary, which is located at the 1-cycle of the O-plane. This gives rise to a two-sphere on $K3$ which is a non-trivial 2-cycle since the corresponding O-plane 1-cycle was non-trivial. Clearly, this 2-cycle is odd under the involution on $K3$ since the involution changes the orientation of the 2-cycle. When we add the fibre, each combination of this 2-cycle with a 1-cycle in the fibre gives rise to a 3-cycle in the threefold. Hence, we get twice as many 3-cycles on the threefold as there are 1-cycles on the O-plane, i.e.\ $2g$ many. This construction is a further motivation
for the cycles that were constructed locally in Section \ref{recO}. If one builds the double cover of $\C^2$ branched along the vanishing
locus of an equation of the form \eqref{2D} one finds the space $\C^2/Z_2$ blown up at the origin. The exceptional cycle of this blow-up is
an odd 2-cycle under the orientifold projection. Its image under the orientifold projection yields precisely the base part of the 3-cycle that controls the O-plane motion in F-theory.

Let us now turn to the contribution coming from $H_2(B,\mathbb{Z})$ and consider a 2-cycle of $B$. Recall that $H_2(B,\mathbb{Z})=H_{2\,+}(K3,\mathbb{Z})$.
Since we assumed $k=0$, we have $H_{2\,+}(K3,\mathbb{Z})^*/H_{2\,+}(K3,\mathbb{Z}) = \mathbb{Z}_2^a$, cf.\ Appendix \ref{nikulinClassification}, and every basis 2-cycle of $H_{2\,+}(K3,\mathbb{Z})$ can be understood as the sum of two (maybe intersecting) basis 2-cycles of $H_{2}(K3,\mathbb{Z})$ that are exchanged by the involution. Then, the difference of these basis 2-cycles gives an element in $H_{2\,-}(K3,\mathbb{Z})$, which can be combined on $K3\times T^2$ with one of the fibre 1-cycles to build an even 3-cycle that descends to the threefold. By this we obtain two 3-cycles of the threefold for each 2-cycle in B. This explains the second contribution in \eqref{conjecture}. For $k=0$, we have $a= b_2(B)$.

Now let us discuss the case of nonzero $k$. This means that we now additionally have $k$ non-trivial rigid two-spheres in $B$ that are part of the fix point locus, i.e.\ which are filled out by the O-plane. Let us consider one of them. Clearly, this cycle has only one pre-image in $K3$, which is left fixed by and thus even under the involution map. Furthermore, we can write \eqref{Nikulin} as
\begin{equation}
 b_2^+(K3)=r=a+2k
\end{equation}
so that it follows that there must be a second even cycle for any fixed $S^2$.
All of these cycles do not lead to any 3-cycle on the threefold and do not contribute to \eqref{conjecture}.
As the quantities $r,a,g$ and $k$ are actually not independent, but related by \eqref{Nikulin}, we find  \eqref{relation}.

Note that from this discussion we see that we can decompose the second cohomology class $H_2(K3,\mathbb{Z})$ of $K3$ into three parts, corresponding to
\begin{itemize}
\item $k$ spheres consisting of fix points plus $k$ further cycles, all of them being even under the involution,
\item $2a$ pairs of cycles which are interchanged
\item $2g$ spheres which are invariant up to an orientation reversal.
\end{itemize}
The fact that these cycles give the correct number of 2-cycles of $K3$, i.e.\
\begin{equation}
 2k + 2a + 2g = 22 \
\end{equation}
follows directly from \eqref{Nikulin}.

\section{D7-branes without obstructions}\label{D7-branes without obstructions}

\subsection{Pulling a single D-brane off the orientifold plane}\label{1doffO}

In this section we now want to leave the orientifold point by moving one D-brane off the O-plane.
The most general form of the hypersurface $\cal D$ which is the position of the D-branes is given by~\cite{Sen:1997bp}
\begin{equation}\label{genald7}
{\cal D}: \quad \eta^2+12 h \chi = 0 \ ,
\end{equation}
where $h,\eta$ and $\chi$ are sections in $[-2K_B], [-4K_B]$ and $[-6K_B]$, respectively.
We will call a D7-brane described by an equation of the form above a \emph{generic allowed} D7-brane.
Note that the equation $h=0$ describes the position of the O-plane $O$. At the orientifold point, the O-plane
coincides with four D-branes, so that $\eta=h^2$ and $\chi=h^3$ and Eq.~\eqref{genald7} reads
\begin{equation}
{\cal D}: \quad h^4 = 0 \ .
\end{equation}
We can now vary the sections $\eta$ and $\chi$ in order to deform the D-branes. Choosing $\eta = h^2$
and $\chi=\tfrac{1}{12} h^2p$, where $p$ is a section in $[-2K_B]$, then yields
\begin{eqnarray}
{\cal D}: \quad  h^3(h+p)= 0 \ .
\end{eqnarray}
The surface $\cal{D}$ consists of four components: three D-branes still coincide
with $O$ while one is deformed and thus separated from the O-plane.
The deformation is given by the generic section $p$ which is of the same degree as $h$.
This fits nicely with the fact that infinitesimal deformations of a surface
correspond to sections in the normal bundle of $O \subset B$.

Let us first return to the example of $\C P^2$. We can count the number of deformations
of the single D-brane that is moved off the O-plane by counting the monomials
of $p$ and subtracting the one complex degree of freedom of overall rescalings.
Note that fixing the O-plane in $B$ generically breaks the
automorphism group of $B$ completely and thus its dimension does not reduce the
number of degrees of freedom. In the present case, $p$ will be a homogeneous polynomial of degree
6 yielding $\frac{1}{2}(6+1)(7+1)-1=27$ complex degrees of freedom.

The number of deformations can also be obtained by analysing sections in the normal bundle of the
D-brane. As the D-brane we are considering is linearly equivalent to $[-2K_B]$, the analysis
of the previous section leading to Eq.~\eqref{moduliRiemann} applies. As the genus of the
D7-brane is given by $10$ in the case of $B=\C P^2$, we can immediately confirm that the
number of deformations is given by $27$. Following an argument similar
to the one presented in Section \ref{fthpers} we thus expect to find $54$ 2-cycles that govern the
displacement of a single D-brane that is equivalent to $[-2K_B]$ from the O-plane in $\C P^2$.
It is clear that a similar computation can be performed for other base spaces.

\subsubsection{3-cycles between O-plane and D-brane}
\label{3cycfrom1cycDbrane}

\begin{figure}
\begin{center}
\includegraphics[height=3cm]{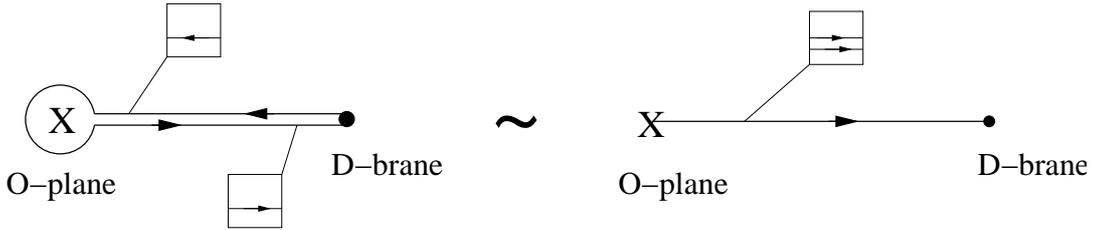}
\end{center}
\caption{\textsl{A possible representative of cycles determining the distance
between O-plane and D-brane in the case of F-theory compactified on $K3$ is given
by a loop that starts and ends at a D-brane and encircles the O-plane. When deforming it,
it looks line a line connecting the D-brane to the O-plane that has twice the horizontal cycle
of the fibre torus as its component in the $T^2$-fibre.}} \label{K3cycle}
\end{figure}

We now construct the 3-cycles that describe the process of moving
a single D-brane off the O-plane. Let us first discuss the analogue of these cycles for F-theory
compactified on $K3$, where O-plane and D-brane are points rather than complex lines.
We can link the two by a path that begins at the D-brane, encircles the O-plane and then ends at
the same D-brane. To construct a 2-cycle, we add the horizontal fibre to every point of this curve,
see Figure~\ref{K3cycle}.

The existence of this cycle can also be demonstrated by the following argument:
two D-branes in the vicinity of an O-plane can be connected by a cycle in two ways: the cycle can pass the O-plane on one
side or the other \cite{Braun:2008ua}. We can find a cycle that connects just one of the two D-branes to the
O-plane by forming the sum (or difference) of these two cycles. The resulting cycle has self-intersection
number $-4$ and can be deformed to any of the two representatives discussed in Figure~\ref{K3cycle}.

Coming back to F-theory compactified on an elliptically fibred Calabi-Yau threefold, we can
generalize this construction as follows. We choose a 1-cycle $A \in H_1({\cal D})$ of the
D-brane and a representative ${\cal A }\subset {\cal D}$. Since we are considering the
case in which the D-brane has the same topology as the O-plane, we can find a
corresponding cycle and representative on the O-plane that coincides with $\cal A$
when D-brane and O-plane are on top of each other. Now we apply the above construction to every point
$p \in {\cal A}$. In other words, we fiber $\cal A$ with the 2-cycles of Figure \ref{K3cycle}.
In this way we obtain $2g({\cal D})$ 3-cycles that measures the distance between the
D-brane and the O-plane.

\subsubsection{3-cycles from intersections between D-brane and O-plane}
\label{cycintd7o7}

\begin{figure}
\begin{center}
\includegraphics[height=3cm]{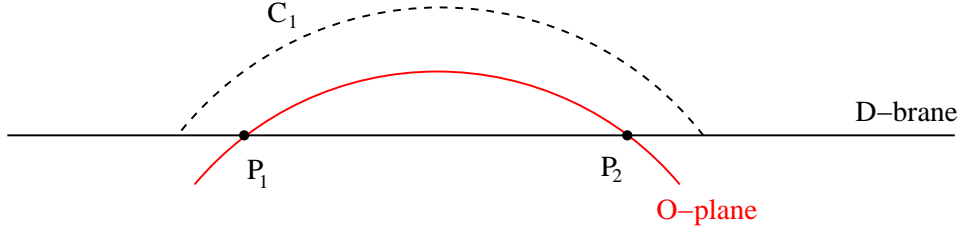}
\end{center}
\caption{\textsl{A 2-dimensional cut through the relative 2-cycle
$C_1$ of type II. Note that $C$ is in fact a half sphere surrounding the
intersections $P_1$ and $P_2$.}} \label{halfspherecycle}
\end{figure}

Another type of cycle can be constructed as follows. Consider two intersection points $P_1$ and $P_2$
of the D-brane with the O-plane (see Figure~\ref{halfspherecycle}). Since $\cal D$ is connected, we can find a loop
$l \subset {\cal D}$ that surrounds both intersection points. This immediately
implies that $l$ can only be contracted if $P_1$ coincides with $P_2$.
We know from Section~\ref{localConstruction} that the disc in B the boundary of which is $l$ can be lifted to a 3-cycle in the threefold $Z$. Indeed, we again fiber the 1-cycle of the torus that degenerates at the D-brane over the disk. This 3-cycle cannot be contracted due to the presence of the O-plane. The involution on the fibre which is part of the monodromy of the O-plane prevents the disk from passing through the O-plane position -- the fibre will simply be ill-defined if the disk intersects the O-plane. Clearly, this 3-cycle has again the topology of a three-sphere and its volume is proportional to the distance between the intersection points with the O-plane.
We can understand this 3-cycle also in another way. We can fiber the $K3$ 2-cycle of Figure~\ref{K3cycle} over the line on the D-brane that connects the two intersection points with the O-plane.

\begin{figure}
\begin{center}
\includegraphics[height=3cm]{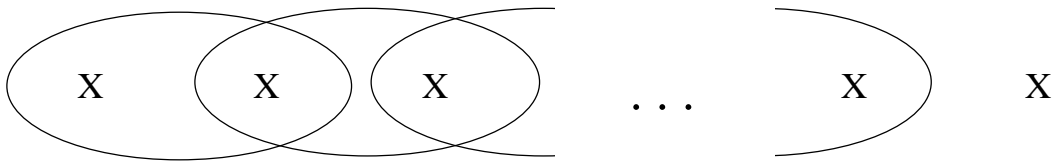}
\end{center}
\caption{\textsl{Boundaries $\partial C_i$ and intersection points with th O-plane $X$ in ${\cal D}$.}}
\label{intersections}
\end{figure}

Let us now count the number of independent cycles that can be constructed in this way.
Suppose there are $I$ intersection points $P_i$ on ${\cal D}$. Let $C_i$ be a disc such that $\partial C_i=l_i$ is a loop
on ${\cal D}$ which surrounds the intersection points $P_i$ and $P_{i+1}$. The boundaries $\partial C_i$
are elements of the first homology group of the D-brane with the O-plane cut out, $H_1({\cal D}\setminus {\cal D}\cap O)$.
Note that $C_i$ and $C_{i+1}$ will generically intersect only in (two) points that are located on the D-brane world volume since
the D-brane and the cycles have codimension two (cf.\ Figure \ref{intersections}). This yields $I$ such
loops\footnote{We identify $i=I+1$ with $i=1$.} and each loop gives a relative 2-cycle $C_i$. The $I$ cycles we construct in this
way are not linearly independent. We can construct the union
\begin{equation}
C = \bigcup_{i=0}^{\frac{I-4}{2}}C_{2i+1} \ .
\end{equation}

The boundary $\partial C$ of the relative 2-cycle $C$ surrounds all intersection points on $\cal D$ except for $P_{I-1}$ and $P_{I}$.
Since D is compact, this is equivalent to saying that $\partial C$ surrounds just $P_{I-1}$ and $P_I$.
Thus, $C$ is relatively homologous to $C_{I-1}$ and hence $C_{I-1}$ is not independent of the others.
In the previous argument we just used half of the $C_i$, namely those where $i$ is odd. We showed that one cycle can be expressed as
a linear combination of the others. The same argument goes through for the
complementary subset of $C_i$ where $i$ is even. Having constructed $I$ 2-cycles $C_i$, we are now left with $I-2$ independent F-theory 3-cycles. This is illustrated in Fig.~\ref{intersections}.

Putting everything together, the number of 3-cycles we obtain by the constructions presented is $2g({\cal D})+I-2$.
As discussed before, the intersection number $I$ of D-brane and O-plane is
nothing but the self-intersection number of $O$. From Eq.~\eqref{selfintersection} we
thus know that $I=4g(O)-4$. Adding the $2g(O)$ cycles coming from the 1-cycles of the D-brane and the
$I-2=4g(O)-6$ cycles coming from the intersections with the O-plane, we arrive at the total number of
$2g(O)+I-2=2(3g(O)-3)$. These cycles determine the infinitesimal separation of the D-brane from the O-plane.
This fits exactly with the number of general deformations of Riemann surfaces obtained before. We conclude that
the union of both kinds of 3-cycles parameterize the location of the D-brane.

Coming back to the example of $B=\C P^2$, we find that $g({\cal D})=10$ and $I=36$,
so that we can construct $54$ cycles. This precisely fits the expectation expressed
at the beginning of this section.

\subsection{More general configurations}

We now want to generalized the discussion to the case of multiple D-branes separated
from the O-plane. When we move multiple branes off the O-plane and let them recombine,
we can no longer describe the resulting D-brane locus in terms of sections in the normal bundle
of the O-plane. Furthermore the worldvolume of the D-brane will in general be no longer be a
generic hypersurface as it is forced to have double intersections with the
O-plane \cite{Braun:2008ua,Collinucci:2008pf}. Namely, $\cal D$ is given by Eq.~\eqref{genald7}.
However, the D-brane curve will still be a generic hypersurface if we consider
one of the following \mbox{configurations}:
\begin{itemize}
\item[i)]{We leave one D-brane on the O-plane. This corresponds to choosing $\eta
= hp$, where $p$ is a section in $[-2K_B]$. This yields
\begin{equation}
{\cal D}: \quad h\underbrace{(hp^2+12\chi)}_{=\chi'} = 0.
\end{equation}
Since $\chi$ is generic, $\chi'$ is. Note that $\chi'$ is a section in
$[-6K_B]$ and therefore describes three recombined D-branes
generically not coinciding with the O-plane. The resulting D-brane locus is
appropriately described by a generic hypersurface in $B$, the zero locus of
$\chi'$, as required. The polynomial $h$ shows that one D-branes still sits on
the O-plane.}

\item[ii)]{We move all D-branes in stacks of two. This case corresponds to the
choice $\chi=0$ so that the D-brane is described by $\eta^2 = 0$ and hence a
generic section in $[-4K_B]$.}
\end{itemize}

In this section we analyse configurations in which the D-brane degrees of
freedom are associated to the deformation moduli of a generic hypersurface of
lower degree. For the whole analysis, we completely fix the O-plane and focus
on the D-brane degrees of freedom.

We start again with the example where $\C P^2$ is the
base space and analyse the general case afterwards. Assume that in addition to the
O-plane we have a D-brane that is linearly equivalent to $[-2mK_B]$, i.e. whose locus in
$B$ is described by the vanishing of a homogeneous polynomial of degree $n=6m$. Due to the D7
tadpole cancellation condition, this means that $4-m$ D-branes still coincide with the O-plane. As the
presence of the O-plane generically breaks the whole automorphism group of $\C P^2$, this
D7-brane has
\begin{equation}\label{degd7}
\tfrac{1}{2}(n+1)(n+2)-1
\end{equation}
complex degrees of freedom.

In the last section we found that a D7-brane that is equivalent to $[-2K_B]$ has $g({\cal D})+\frac{I}{2}-1$ complex degrees of freedom and gives rise to $2g({\cal D})+I-2$ 3-cycles in the threefold. In the remainder of this section, we show that this statement holds for any D7-brane that is equivalent to $[-2mK_B]$ for any $m$. The reader primarily interested in results may wish to skip the rest of this section and continue with Section~\ref{moreDoffO}.

Let us now try to find an analog of Eq.~\eqref{degd7} for an arbitrary base
space $B$. We will formulate all relevant quantities in terms of the
self-intersection of the anti-canonical divisor of the base, $(-K_B)\cdot (-K_B)=S_B$. For any
F-theory model, the worldvolume of the O-plane is equivalent to the divisor $-2K_B$. We consider
the situation with $m$ recombined D-branes $\cal D$ and $4-m$ D-branes coinciding with
the O-plane. In order to apply the Riemann-Roch Theorem, cf.\ Eq.~\eqref{dimHolSections},
we need to know the degree of the canonical divisor of $\cal D$, denoted by $K_{\cal D}$.
It is given by
\begin{equation}
\deg K_{\cal D} = 2g({\cal D})-2.
\end{equation}
The Euler characteristic of $\cal D$ is
\begin{equation}
\chi({\cal D})=\left[-K_B+2mK_B\right]\cdot\left[-2mK_B\right]=-2m(2m-1)S_B \ ,
\end{equation}
so that the genus is given by
\begin{equation}
g({\cal D})=m(2m-1)S_B+1 \ .
\end{equation}
Thus we find the degree of $K_{\cal D}$ to be
\begin{equation}
\deg K_{\cal D} =2m(2m-1)S_B \ .
\end{equation}
The self-intersection number of $\cal D$ is $\deg
N_{\cal D} = (2m)^2S_B$, so that the condition $\deg K_{\cal D} < \deg
N_{\cal D}$ is satisfied. Hence we can use Eq.~\eqref{dimHolSections} to find
the number of valid deformations:
\begin{equation}\label{genmatch}
\begin{aligned}
h^0(N_{\cal D}) &= (2m)^2S_B -
m(2m-1)S_B =
m(2m+1)S_B \\
&=1+m(2m-1)S_B+2m S_B-1=g({\cal D}) + \frac{I}{2}-1 \ .
\end{aligned}
\end{equation}
In the last line we have used that the number of intersections between the D-brane and
the O-plane is
\begin{equation}
I=\#\left(O \cap {\cal D}\right) =
(-2K_B)\cdot(-2mK_B)=4m S_B \ .
\end{equation}
We thus expect to find $2g+I-2$ 3-cycles that govern the deformation of the D-brane locus.

From the relation derived in the last paragraph it is clear how the 3-cycles that
control the motion of a D-brane arise:
On the one hand, we can build a 3-cycle from every 1-cycle
of the D-brane, using the construction given in Section~\ref{recomD}.
On the other hand, we can build cycles that measure the distance between intersections
of the D-brane with the O-plane, as discussed in Section~\ref{3cycfrom1cycDbrane}.

This can also be understood from the perspective of the $K3$ double cover as we now
explain qualitatively. It is known that for a smooth D-brane in the double cover, i.e. one that
does not have double intersections with the O-plane, the deformations are given by
1-cycles of the D-brane that are odd under the involution~\cite{Jockers:2004yj}.
As this is the situation discussed in this section, we should be able to link
the 3-cycles we have constructed to odd 1-cycles of the D-brane in the
double cover.

A 1-cycle of a D-brane ${\cal D}$ in $B$ that has been moved off the O-plane
generically does not intersect the O-plane. Furthermore, we can always deform the 1-cycle such that its winding number is zero with respect to the O-plane. 
Therefore, this
1-cycle has two pre-images in the double cover $K3$ which are interchanged by the involution.
The sum of both is even under the involution and therefore descends to the 1-cycle of ${\cal D}$
we started with. The difference of both, however, is odd under the projection and should refer to a
deformation of the D-brane. This suggests a fact already discussed: 1-cycles of ${\cal D}$ are related to 3-cycles
of the Calabi-Yau threefold. Furthermore, we can consider a line on ${\cal D}$ connecting two intersection
points of ${\cal D}$ with the O-plane and go to the double cover $K3$.  This line then becomes two lines joined at their
end points, i.e.\ a non-trivial closed 1-cycle on the double cover D-brane that is odd under the involution. Thus,
there should be a second kind of 3-cycles which are closely related to the intersections of the D-brane with the O-plane,
supporting our claim that we can construct 3-cycles from intersections between the D-brane and the O-plane.

As the intersections between O-plane and D-brane are points on a complex curve, one naively expects each intersection
point to correspond to a complex degree of freedom. From the relation between moduli and 3-cycles it follows that there
should roughly be $2I$ 3-cycles that stem from intersection points between D-brane and O-plane. This is, however, not the
case, as we only found half of that. As we explain in Appendix~\ref{abelSection}, a more detailed analysis using Abel's 
theorem reveals constraints which reduce the number of deformations of the intersection points.

Compared to Section~\ref{1doffO}, our cycle analysis is complicated by the fact that the O-plane can in principle pierce a disc that ends
on a 1-cycle.\footnote{As we will discuss in more detail later, this is ultimately related to the structure of
branch cuts on the D-brane when building the double cover.} The monodromy of the O-plane then prevents the construction
of a 3-cycle as its fibre part is transformed to a different cycle upon encircling the O-plane locus. To tell if we can find a disc that ends on a given loop
on the D-brane and does not intersect the O-plane, we need to check that the winding number of this loop
around the O-plane vanishes.
We can define the winding number on the first homology of the D7-brane with the intersection points with the
O-plane cut out, $H_1({\cal D},O \cap {\cal D})$, and then project to the subspace of zero winding number. 
Since $H_1({\cal D},O \cap {\cal D})$ has the dimension $2g+I-1$ and there are elements of $H_1({\cal D},O \cap {\cal D})$ that have a non-zero winding number, this projection leads to a subspace of dimension $2g+I-2$, i.e.\ we find $2g+I-2$ independent cycles we can use to construct non-trivial 3-cycles of the elliptic fibration. This reproduces the
result that is expected from an analysis of the degrees of freedom of a D-brane.

Let us again come back to the example of $\C P^2$. In this case one easily finds the
numbers $g({\cal D})=(n-1)(n-2)/2$ and $I=6n$, so that
\begin{equation}
\underbrace{\frac{1}{2}(n-1)(n-2)}_{=g({\cal
D})}+\underbrace{3n-1}_{=\frac{I}{2} -1}=\underbrace{\frac{1}{2}(n+1)(n+2)-1}_{
= \dim_\C \text{Def}({\cal D})},\label{matchcp2}
\end{equation}
which exactly matches the number of degrees of freedom, as given by \eqref{degd7}.

Note that we can use the reasoning presented in this section also for the
situation discussed at the beginning of this section, in which a
single D-brane is moved off the O-plane. Although we used different
3-cycles in both cases, the results agree as expected. The two sets of
cycles just give a different basis of the third homology group of the threefold.

We can also give an inductive construction of 3-cycles in terms of (relative) 1-cycles
as long as all D-branes are described by completely generic hypersurfaces. We start with
the case in which the D-brane locus and the O-plane locus coincide. These cycles were
discussed at length in Section~\ref{DonO}. New cycles appear when the first D-brane is moved
away from the O-plane, namely the cycles given in Section~\ref{3cycfrom1cycDbrane}. We used
in this analysis that the D-brane is given by a section in the normal bundle of the O-plane.

We can now independently move two D-branes
off the O-plane, both given by sections in the normal bundle of $O$. Additionally to the
cycles for described in Section~\ref{recomD}, there are intersections
between the two D-branes. Thus, the D-brane locus is a nodal Riemann surface $\overline{\cal D}$
with $I_{DD}$ nodes, where $I_{DD}$ denotes the number of intersection points of the two D-branes.
By generic deformations of these singular intersection points, the D-branes recombine
at these nodes as described in Section~\ref{recomD}, yielding a smooth Riemann surface $\cal D$. Note
that the genus $g({\cal D})$ of $\cal D$ is identical to the arithmetic
genus $p(\overline{\cal D})$ of $\overline{\cal D}$. For the arithmetic genus
the following identity holds~\cite{Hori:2003ic}
\begin{equation}
p(\Sigma) = \delta+1+\sum_{i=1}^{k}(g_i-1), \label{arithmeticgenus}
\end{equation}
where $\Sigma$ is the nodal Riemann surface with $\delta$ nodes and $k$
irreducible components which have the geometric genus $g_i$, respectively. In
our case of interest, Eq.~\eqref{arithmeticgenus} reduces to
\begin{equation}
g({\cal D})=p(\overline{\cal D})=2g({\cal D}_0)+I_{DD}-1,
\end{equation}
where ${\cal D}_0$ denotes the single D-brane. Using Eq.~\eqref{genmatch} we can
immediately give an expression for the number of independent deformations of
$\cal D$:
\begin{eqnarray}
\dim_\C \text{Def}({\cal D})&=&h^0(N_{\cal D})=g({\cal D})+\frac{I_{OD}}{2}-1
\nonumber \\
&=& 2g({\cal D}_0)+I_{DD}-2+I_{OD_0} \nonumber \\
&=& 2\dim_\C \text{Def}({\cal D}_0)+I_{DD}. \label{recomDbrane}
\end{eqnarray}
Here $I_{OD_0}$ denotes the intersection number between $O$ and ${\cal D}_0$
and we used that $I_{OD}=2I_{OD_0}$. The first term in the last line in
Eq.~\eqref{recomDbrane} are the degrees of freedom obtained
by moving the D-branes ${\cal D}_0$ independently. The second part, namely
$I_{DD}$, gives the number of recombination parameters. Each
intersection point gives exactly two cycles. In fact, these are
locally the recombination cycles obtained in Section~\ref{recomD}.
In the same way we can discuss the case of three D-branes moved off the O-plane.

\section{D7-branes with obstructions} \label{moreDoffO}

\subsection{D-brane obstructions} \label{Obstructions}
\begin{figure}
\begin{center}
\includegraphics[height=2.5cm]{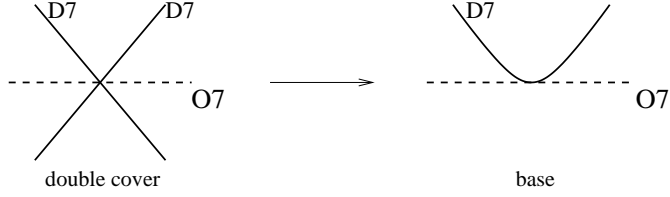}
\end{center}
\caption{\textsl{A situation in which two D7-branes intersect an O7-plane
in the same point produces a D7-brane touching the O-plane after modding out
the orientifold action and squaring the coordinate transverse to the O7-plane.
}}\label{2btof}
\end{figure}

In the weak coupling limit the D7-brane locus is not given by the zeros
of a generic polynomial, but by the zeros of a polynomial of the form
\begin{equation}\label{obstructed_Weierstrass}
 {\cal D}: \quad \eta^2+12 h \chi = 0 \ .
\end{equation}
We refer to D7-branes that are described by an equation of this form as \emph{generic allowed}
D7-branes. As has recently been discussed \cite{Braun:2008ua,Collinucci:2008pf},
this form forces the D7-brane to have double intersections
with the O7-plane. From the perspective of F-theory this means that the D7-brane
forms a parabola touching the O-plane in the origin, see Figure~\ref{2btof}.

Let us try to understand this configuration from the double cover perspective.
Consider two D-branes at $x=\pm z$ and the involution $z \to - z$, which fixes the O-plane at $z=0$.
After modding out the involution, our space looks locally
like the upper half plane. In order to make contact with the F-theory picture,
we introduce a new coordinate $\tilde{z}=z^2$ and find that this situation is
described by an O-plane at $\tilde{z}=0$ and a D-brane at $\tilde{z}=x^2$.
Note that a single D7-O7 intersection in F-theory (which does not occur in the weak coupling limit),
corresponds to a single D7-brane that is mapped onto itself by the
orientifold projection. This configuration, where the D-brane sits e.g.\ at $x=0$ is allowed
in the presence of a second D-brane that coincides with the O-plane.

In \cite{Braun:2008ua} it was observed that the difference between the degrees
of freedom of a generic allowed D7-brane and the degrees of freedom of a
generic hypersurface of the same degree is given by half the number of intersections
between the D7-brane and the O7-plane\footnote{Here we of course count the topological
intersections between two generic surfaces that are homologous to the D7-brane and the O7-plane.}.
We checked this explicitly only for base spaces $\C P^2$ and $\C P^1 \times \C P^1$. Here we
extend this analysis to all possible base spaces. We use the fact that a generic hypersurface
${\cal D}_\textrm{gen}$ that is linearly equivalent to $2mK_B$ has
\begin{equation}
\dim_\C \text{Def}({\cal D}_\textrm{gen})=m(2m+1)K_B\cdot K_B=m(2m+1)S_B \label{defhyp}
\end{equation}
deformations. To simplify equations we again use $S_B$ as a shorthand for $(-K_B)\cdot (-K_B)=S_B$.
As in the last section we keep the O-plane fixed so that the automorphism group of the base is completely
broken. As the number of double intersections is given by $2mS_B$, the double intersections lead
to $2mS_B$ constraints, so that we expect the number of deformations encoded in Eq.~\eqref{genald7}
to be
\begin{equation}\label{def_obstructedbrane}
\dim_\C \text{Def}({\cal D}) = m(2m+1)S_B-2mS_B=m(2m-1)S_B\ .
\end{equation}

The degrees of freedom in the expression (\ref{genald7}) are given by the
number of monomials in $\eta$ and $\chi$, minus an overall rescaling
and the redundancy that corresponds to shifting $\eta$ by $h\alpha$, $\alpha$ being
a polynomial of appropriate degree. To compute the number of monomials, we note that
we can take the polynomials $\eta$, $\chi$ and $\alpha$ to define hypersurfaces on
their own and compute the number of their deformations using Eq.~\eqref{defhyp}.
The number of monomials is then given by the number of deformation plus one. For a D7-brane that is
equivalent to $-2mK_B$, $\eta$, $\chi$ and $\alpha$ are sections of $[-mK_B]$,
$[-(2m-2)K_B]$ and $[-(m-2)K_B]$, respectively. Thus the degrees of freedom in Eq.~\eqref{genald7} are given in this case by
\begin{equation}\label{def_obstructedbrane2}
\begin{aligned}
\dim_\C \text{Def}({\cal D}) & = \underbrace{\tfrac{1}{2}m(m+1)S_B+1}_{M(\eta)}+\underbrace{(m-1)(2m-1)S_B+1}
_{M(\chi)}
-\underbrace{\left(\tfrac{1}{2}(m-2)(m-1)S_B+1\right)}_{M(\alpha)}-1
\\ &=m(2m-1)S_B \ ,
\end{aligned}
\end{equation}
which coincides with \eqref{def_obstructedbrane}. 

For a generic hypersurface ${\cal D}_\textrm{gen}$ one can actually show that in the double cover its deformation space is isomorphic to $H^{(1,0)}_-({\cal D}'_\textrm{gen})$, where ${\cal D}'_\textrm{gen}$ is the corresponding preimage of ${\cal D}_\textrm{gen}$ in the double cover \cite{Jockers:2004yj}:
\begin{equation}\label{H(1_0)_generic}
\begin{aligned}
h^{(1,0)}_-({\cal D}'_\textrm{gen})&=g({\cal D}_\textrm{gen})+\tfrac{1}{2}I_{{\cal D}_\textrm{gen}-O7}-1  \\
&=m(2m-1)S_B+1+2m S_B-1  \\
&=m(2m+1)S_B \ ,
\end{aligned}
\end{equation}
where the first of the above equalities follows from the Riemann-Hurwitz theorem \cite{Griffiths:1994}. 
We want to stress that the above statement is not true any more for a generic allowed brane ${\cal D}$ and its double cover ${\cal D}'$. 
More precisely, we see from \eqref{def_obstructedbrane2} that the number of degrees of freedom is exactly $g({\cal D})-1$. A comparison with the computation in \eqref{H(1_0)_generic} suggests that the cycles in $H^{(1,0)}_-({\cal D}')$ which are related to the double intersections do not give rise to deformations of ${\cal D}$. 

Let us now perform an analogous computation for the generic allowed brane
${\cal D}$ and its double cover ${\cal D}'$ to confirm this observation.
As we already discussed in Figure~\ref{2btof} and below
\eqref{obstructed_Weierstrass}, ${\cal D}'$ is a smooth brane apart
from its self-intersections (which occur at every intersection
point with the O-plane). Removing these singular points from ${\cal D}'$, 
we obtain a smooth Riemann surface with punctures on which the
$Z_2$ orbifold projection acts freely. Subsequently, we compactify this
punctured Riemann surface in the obvious way, by adding one point per
puncture. The result is a smooth compact Riemann surface with free
$Z_2$ action, which we continue to call ${\cal D}'$ by abuse of notation. 
While this smooth Riemann surface is not realized as a submanifold of the
double-cover Calabi-Yau, its $Z_2$ projection ${\cal D}$ is still our
familiar generically allowed D-brane given as a submanifold of the base
$B$. By standard arguments \cite{Jockers:2004yj}, its allowed deformations
correspond to $Z_2$-odd sections of the canonical bundle $K_{{\cal D}'}$
of ${\cal D}'$, which is understood as a smooth Riemann surface as
explained above. Thus, repeating the calculation of \eqref{H(1_0)_generic},
the number of deformations is given by
\begin{equation}
h^{(1,0)}_-({\cal D}')=g({\cal D})-1 =m(2m-1)S_B \ ,
\end{equation}
where we have again used the Riemann-Hurwitz theorem, but now for a freely
acting involution. This agrees with our previous results. The advantage
of this new derivation is that we are now able to specify which bundle
over ${\cal D}$ encodes these deformations. Indeed, while the even
sections of $K_{{\cal D}'}$ correspond to sections of $K_{\cal D}$, the
odd sections of $K_{{\cal D}'}$ can be understood as sections of
a `twisted' canonical bundle $\tilde{K}_{\cal D}$ over ${\cal D}$. The
latter is is defined as the $Z_2$ projection of $K_{{\cal D}'}$, where
the $Z_2$ action on ${\cal D}'$ is supplemented by a `$-1$' action on the
fibre. Locally, in a small neighbourhood of an intersection point
with the O-plane, this is still the canonical bundle of ${\cal D}$,
in agreement with the discussion of \cite{Beasley:2008dc,Beasley:2008kw,Cordova:2009fg}.

\subsection{Recombination for double intersection points} \label{RecObstr}
Now we want to understand the number of 3-cycles from the threefold perspective and explain why the number of 3-cycles is reduced by $I$ when compared with our results in Section~\ref{D7-branes without obstructions}. As mentioned above, we need to understand the winding numbers of the D-brane 1-cycles relative to the O-plane. First we discuss this locally for a single 1-cycle and then in Section \ref{ObstructionsCycles} analyze the global situation.

\begin{figure}
\begin{center}
\includegraphics[height=4cm, angle=0]{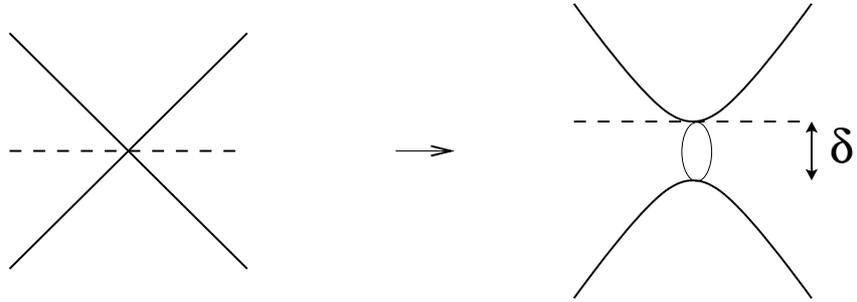}
\end{center}
\caption{\textsl{Recombination of the two D-branes at the common intersection point with the O-plane leads to a double intersection point of the recombined D-brane with the O-plane.}}
\label{recombprocessobstr}
\end{figure}

\begin{figure}
\begin{center}
\includegraphics[height=4cm, angle=0]{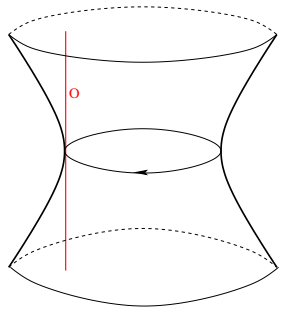}
\end{center}
\caption{\textsl{After recombination, the O-plane has a double intersection point with the recombined D-brane.}}
\label{recombinationobstructed}
\end{figure}

\begin{figure}
\begin{center}
\includegraphics[height=4cm, angle=0]{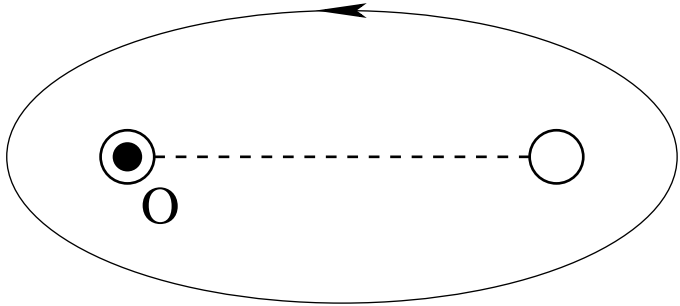}
\end{center}
\caption{\textsl{Recombination at a double intersection point blows up the branch cut between the two branch points. One of the branch points is the point where the O-plane meets the recombined D-brane (illustrated by the black dot). The loop around the branch cut illustrates the non-trivial 1-cycle encircling the throat.}}
\label{branchcutoplane}
\end{figure}

Let us again consider the recombination of two intersecting D-branes, cf.\ Section~\ref{recomD}, but now in the presence of the O-plane at the intersection point. Furthermore, we assume that we have already moved the fourth D-brane off the O-plane such that we describe the D-branes by Eq.~\eqref{genald7}. If we set in the local model
\begin{equation}
\begin{aligned}
 h & =z \ , \\
 \eta & =x \ , \\
 \chi & = \tfrac{1}{12} (z - \delta) \ ,
\end{aligned}
\end{equation}
we have the situation of an O-plane at $z=0$ and a D-brane given at
\begin{equation}\label{brane_recomb_obstr}
 x^2 = z (z-\delta) \ .
\end{equation}
For $\delta=0$, it parameterizes the situation of two D-branes at $x \pm z =0$ and an O-plane at $z=0$, all of them intersecting at the origin, as shown in the left picture in Figure~\ref{recombprocessobstr}. If we now give $\delta$ a non-zero value, we get to the recombined situation to the right in Figure~\ref{recombprocessobstr}.
Here the diameter of the throat connecting the two D-branes is given by $\delta$.
After recombination, the O-plane touches the D-brane tangentially at $x=z=0$, see Fig.~\ref{recombinationobstructed}.
With help of Eq.~\eqref{brane_recomb_obstr}, we can picture the D-brane as the $z$-plane with a branch cut between the two branch points at $z=0$ and $z=\delta$, as shown in Figure~\ref{branchcutoplane}.
The intersection with the O-plane is also given by $z=0$, and therefore, the D-brane 1-cycle encircles the O-plane intersection exactly once. However, since the O-plane is just a plane parameterized by $z=0$, any loop on the D-brane world-volume encircling $z=0$ encircles, when understood as a curve in $B$, the O-plane exactly once, too. Thus, the D-brane 1-cycle has a winding number of one relative to the O-plane.

Since the D-brane 1-cycle wraps the O-plane exactly once, we cannot build a 3-cycle out of it as done in Section~\ref{recomD}.
Because of the involution of the O-plane the fibre would simply be ill-defined. From the double cover perspective, it is clear that
the 3-cycle construction fails for a D-brane 1-cycle with an odd wrapping number:
In the double cover, the lift of this 1-cycle is not closed and does not define a D-brane 1-cycle in the double cover.

\subsection{Threefold cycles, obstructions and the intersection matrix} \label{ObstructionsCycles}

In this section we outline how threefold three-cycles arise from the topology of a generic allowed D-brane, analogously 
to Section~\ref{1doffO}. We do not provide a rigorous proof but rather sketch the construction of threefold 
3-cycles from the 1-cycles of a generic allowed D-brane in the presence of a `naked' O-plane.

\begin{figure}
\begin{center}
\includegraphics[height=8cm, angle=0]{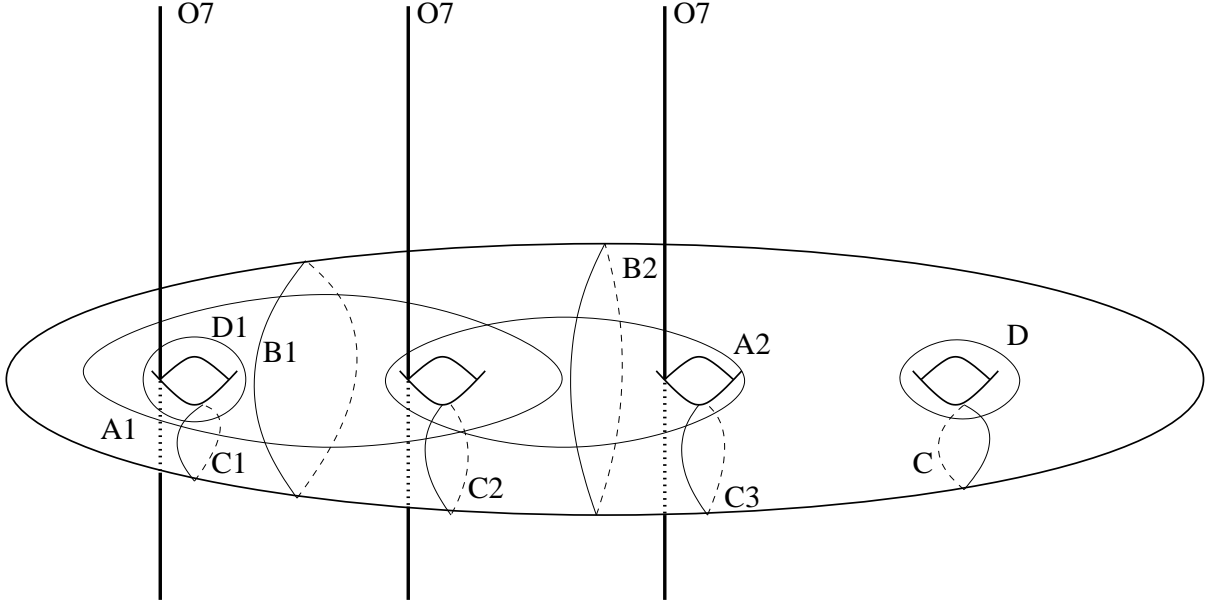}
\end{center}
\caption{\textsl{Here we see the various curves on the D-brane which lead to 3-cycles of the threefold. The curves C and D give rise to 3-cycles in the usual way, as explained in Section~\ref{recomD}. The curves A1 and A2 have an even wrapping number and therefore also lead to non-trivial 3-cycles. The dual cycles are constructed from the curves C1, C2 and C3 and are linearly equivalent to the 3-cycles over B1 and B2, which are non-trivial due to the O-plane monodromy.}}
\label{CyclesObstructions}
\end{figure}

In this section we want to discuss the threefold 3-cycles in the case of a generic allowed D-brane and a `naked' O-plane.
Using \eqref{genmatch} we can express the number of degrees
of freedom of a generic allowed brane in terms of its genus:
\begin{equation}
\dim_\C \text{Def}({\cal D})=m(2m-1)S_B -1=g({\cal D})-1.
\end{equation}
We thus expect that we can construct $2g-2$ 3-cycles.
All intersections of the D-brane with the O-plane are double intersections, which makes it impossible to construct cycles in the spirit of Section~\ref{cycintd7o7}. However, we still can build 3-cycles related to the D-brane 1-cycles, as we explain now. For this, we first choose a symplectic basis for the 1-cycles of the D-brane and then discuss how the basis 1-cycles lead to 3-cycles in the threefold, depending on the wrapping number of these 1-cycles with respect to the O-plane.
We illustrate the curves on the D-brane that lead to 3-cycles in Fig.~\ref{CyclesObstructions}.
Note that Fig.~\ref{CyclesObstructions} represents a local picture of the D-brane, ignoring the topology of $B$ and of the O-plane. 

Consider a symplectic pair of D-brane 1-cycles.
First assume that both 1-cycles have zero wrapping number with respect to the O-plane. If this occurs, we can construct a 3-cycle over each of them as explained in Section~\ref{recomD}, giving a symplectic pair of 3-cycles which has no intersection with any of the other 3-cycles. An example of such a symplectic pair is given by C and D in Fig.~\ref{CyclesObstructions}.

Next consider a symplectic pair of 1-cycles where one of the 1-cycles has wrapping number one while the paired cycle has still wrapping number zero. An example of such a pair is given by C1 and D1 in Fig.~\ref{CyclesObstructions}, where D1 has wrapping number one and C1 has wrapping number zero.
If we try to use the same technique as in Section~\ref{recomD} and fibre the usual 1-cycle over a disc ending on D1, we cannot close this 3-cycle since the monodromy of the O-plane inverts the orientation of the fibre 1-cycle. One might have the idea to use a loop corresponding to 2D1, which surrounds the O-plane twice. Since the orientation of the fibre 1-cycle is inverted twice, the corresponding 3-cycle is closed. However, this 3-cycle must be trivial since it is by construction symmetric under orientation reversal, i.e.\ the 3-cycle is equivalent to minus itself. 

This result has several consequences. First of all, we see that if we have a symplectic pair of 1-cycles where both have an even wrapping number, we can always add 2D1 such that the wrapping numbers are zero and apply the construction method of Section~\ref{recomD} to find two 3-cycles. This should be seen in contrast to Section~\ref{D7-branes without obstructions}, where one had to restrict to the D-brane 1-cycles that have zero wrapping number and thereby reduced the number of appropriate 1-cycles by one.
Secondly, the construction of 3-cycles from cycles with odd wrapping number must be modified. 
Note that if both 1-cycles of a symplectic pair have an odd wrapping number, we can replace one of them by the sum of both, leading to a symplectic pair of 1-cycles where only one wrapping number is odd.
Thus, the remaining case is that exactly one of the two D-brane 1-cycles has an odd wrapping number.\footnote{The number of such symplectic pairs is actually $8 S_B$, as we explain now. If we consider Eq.~\eqref{genald7} with $\chi=\tfrac{1}{12} h \alpha^2$, this corresponds to two D-branes at $\eta \pm h \alpha$ which then can be recombined at each of the $8 S_B$ intersection points. With the result of Section~\ref{RecObstr} we conclude that there are $8 S_B$ D-brane 1-cycles with odd wrapping number. These 1-cycles correspond to the $I/2=8 S_B$ double intersections of the D-brane with the O-plane.}

As discussed above, over a D-brane 1-cycle with odd wrapping number no 3-cycle can be defined due to the involution that is part of the O-plane monodromy. However, we can use the sum of two such 1-cycles which is represented by a curve encircling two of the double intersection points with the O-plane.
Examples of such curves are denoted in Fig.~\ref{CyclesObstructions} by ${\rm A}1={\rm D}1-{\rm D}2$ and ${\rm A}2={\rm D}2-{\rm D}3$. 
Note that since 3-cycles constructed out 2D1, 2D2 oder 2D3 are trivial, the third combination ${\rm D}1+{\rm D}3$ does not lead to any independent 3-cycle but to the one constructed out of ${\rm A}1+{\rm A}2$.
Thus, this construction gives us one 3-cycle less than there are D-brane 1-cycles with odd wrapping number.

Now we turn to the 1-cycles that are dual to those with odd wrapping number, represented by C1, C2 and C3 in Fig.~\ref{CyclesObstructions}. Since these 1-cycles have even wrapping number we can construct 3-cycles $C_i$ out of them. However, there is a linear dependence between them, as we show now. Consider the curves B1 and B2 in Fig.~\ref{CyclesObstructions}. They both lead to non-trivial 3-cycles $B_i$ due to the intersection points of the D-brane with the O-plane. Since the involution of the O-plane inverts the orientation of the fibre 1-cycles, the 3-cycle $C^i$ comes back to minus itself when once ``encircling'' the O-plane.
More precisely, if we denote the corresponding 3-cycles by $C_i$ and $B_i$, due to the monodromy of the O-plane there is
\begin{equation}
  2 C_1 = B_1 \ , \quad 2 C_2 = B_2-B_1 \ , \quad  2 C_3 = - B_2 \ ,
\end{equation}
leading to
\begin{equation}
  2 C_1 + 2 C_2 + 2 C_3 = 0 \ .
\end{equation}
On a general D-brane, the linear dependence analogously reads
\begin{equation}
 2 \sum_{i=1}^{I/2} C_i = 0 \ ,
\end{equation}
where the sum runs over all 3-cycles $C_i$ which are coming from D-brane 1-cycles dual to those with odd wrapping number. Thus, we find that the total number of 3-cycles coming from the D-brane sector is two less than the number of 1-cycles of the D-brane, leading to $2g({\cal D})-2 = \chi({\cal D}) = 56 S_B$ 3-cycles. This coincides with the result we achieved in Section~\ref{Obstructions}.

Let us now discuss the intersection matrix of the 3-cycles discussed here. As already stated above, the intersection matrix of the 3-cycles that come from 1-cycles with even wrapping number is the standart symplectic one. For the 3-cycles which are constructed out of curves of type A and B, the corresponding 3-cycles $A_i$ and $\frac{1}{2} B_j = \sum_{i=1}^j C_i$ also form a symplectic basis.\footnote{Note that the $A_i$ do not intersect each other due to the monodromy of the O-plane.} Thus, we find the symplectic basis for the 3-cycles of the threefold that come from the D-brane sector.

\section{Conclusions and outlook}

We have discussed a parameterization of type IIB orientifold and brane moduli in
terms of the complex structure moduli space of the corresponding elliptically
fibred Calabi-Yau threefold. We succeeded to identify the 3-cycles of the Calabi-Yau threefold
and showed they match the number of open string moduli. We were able to apply our analysis
also in the case in which the D-brane is forced to have double intersections with the
O-plane.

A similar analysis was done in a previous paper \cite{Braun:2008ua} in the
case of F-Theory compactifications on $K3$-surfaces, which correspond to
orientifold models on $\C P^1$ in the weak coupling limit. In the present work
we made a first step towards compactifications of higher dimensions. The result
that we have complete control over the brane moduli space in terms of the
complex structure moduli of F-theory even in this more complicated geometrical
setting, feeds on hope that this will also be possible in the case of realistic
compactifications down to four dimensions. Progress in this direction would
yield an important tool for realistic model building, in particular regarding
flux stabilization of configurations with different gauge symmetries in the spirit of \cite{Braun:2008pz}.
Proceeding this way, one will encounter the problem that there is no classification of spaces that
have a double cover Calabi-Yau space, since the Calabi-Yau condition in three
dimensions does not single out exactly one space. On the other hand, Fano threefolds,
which qualify as base spaces of elliptic fourfolds, have been classified, see e.g. \cite{Mori}.
Additionally, one will have to explain the status of `Whitney umbrella'-like
intersections \cite{Collinucci:2008pf} in the context of fourfolds.

\section*{Acknowledgements}
We have greatly benefited from conversations and correspondence with Markus Banagl, Thomas Grimm, Tae-Won Ha, Albrecht Klemm, 
Denis Klevers, Bernd Siebert, Roberto Valandro, Timo Weigand and Rainer Weissauer. A.H. is grateful to the Berkeley Center 
for Theoretical Physics for hospitality. The work of S.G. is partially supported by the Swiss National Science Foundation. 
H.T. is supported by the German Science Foundation (DFG) under the Collaborative Research Center (SFB) 676.
\appendix

\section{Rational surfaces} \label{complexSurfaces}
We know from the classification of Nikulin, reviewed in Appendix \ref{nikulinClassification}, that rational surfaces naturally appear as base spaces for $K3$ orientifolds in type IIB.
Rational surfaces can be
obtained by blowing up $\C P^2$ or Hirzebruch surfaces $\Hirz[n]$ and play an important role in our considerations.\footnote{$\C P^2$ and
$\Hirz[n]$ themselves are called minimal rational surfaces. A surface is called minimal
if it does not contain a curve with self-intersection $(-1)$. If a surface actually does
contain such curves, these curves can be blown down in order to obtain a minimal surface.}
Their main properties are
summarized in this appendix.
In Appendix \ref{delPezzo} we briefly discuss del Pezzo surfaces, which are (except $ \C P^1 \times \C
P^1$) blow-ups of $\C P^2$. In Appendix \ref{Hirzebruchsurfaces} we then give a short review on Hirzebruch surfaces.

\subsection{Del Pezzo surfaces} \label{delPezzo}
A complex two dimensional manifold $Y$ is called a del Pezzo surface if the
anticanonical bundle is positive definite, i.e.\ it has positive intersection
number with every curve in $Y$.

\begin{figure}
\begin{tabular}{c@{\hspace{3cm}}c@{\hspace{3cm}}c}
\includegraphics[height=4cm]{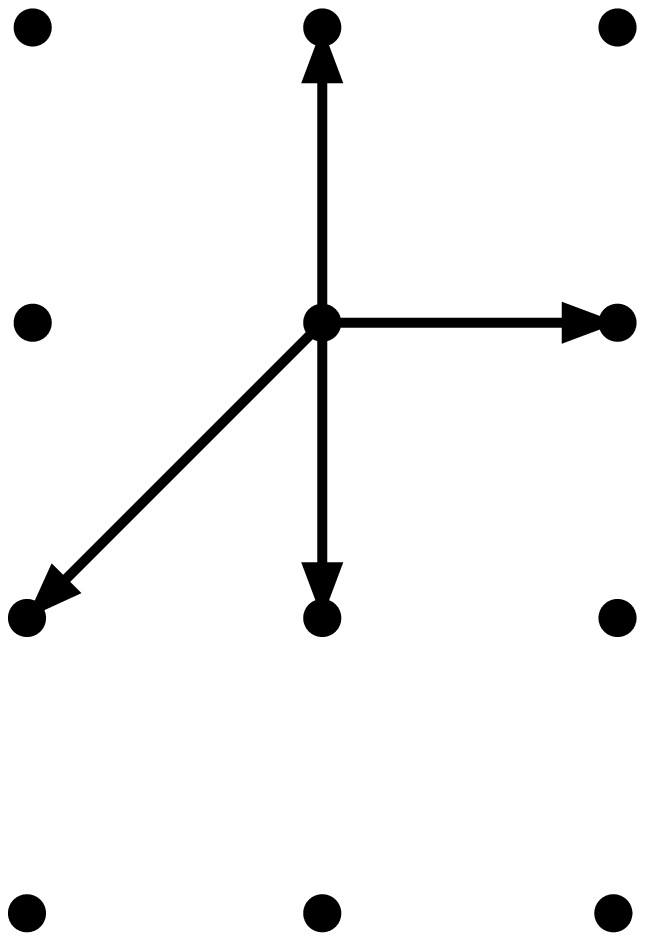}  &
\includegraphics[height=4cm]{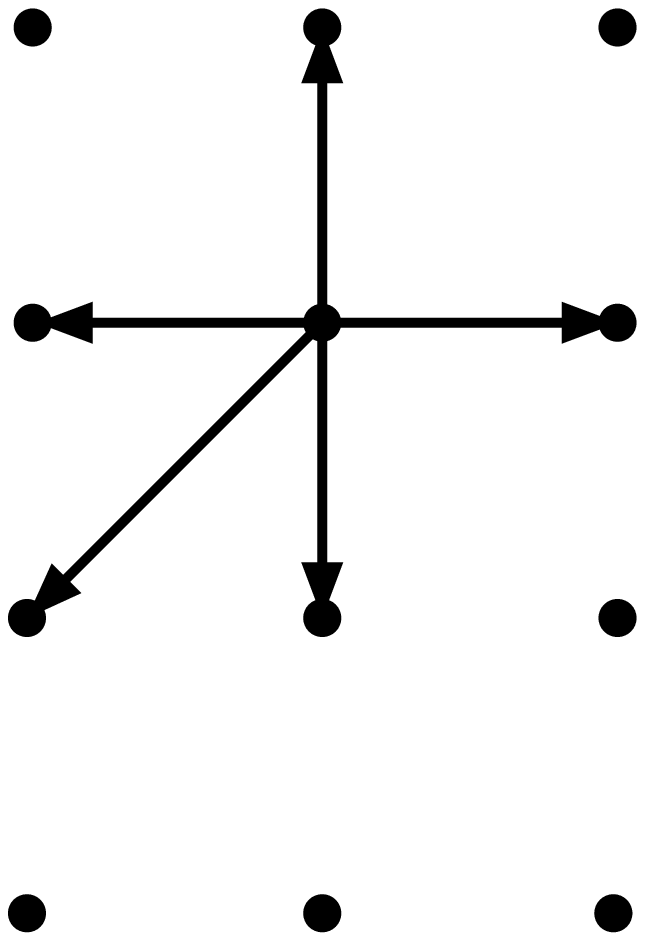}  &
\includegraphics[height=4cm]{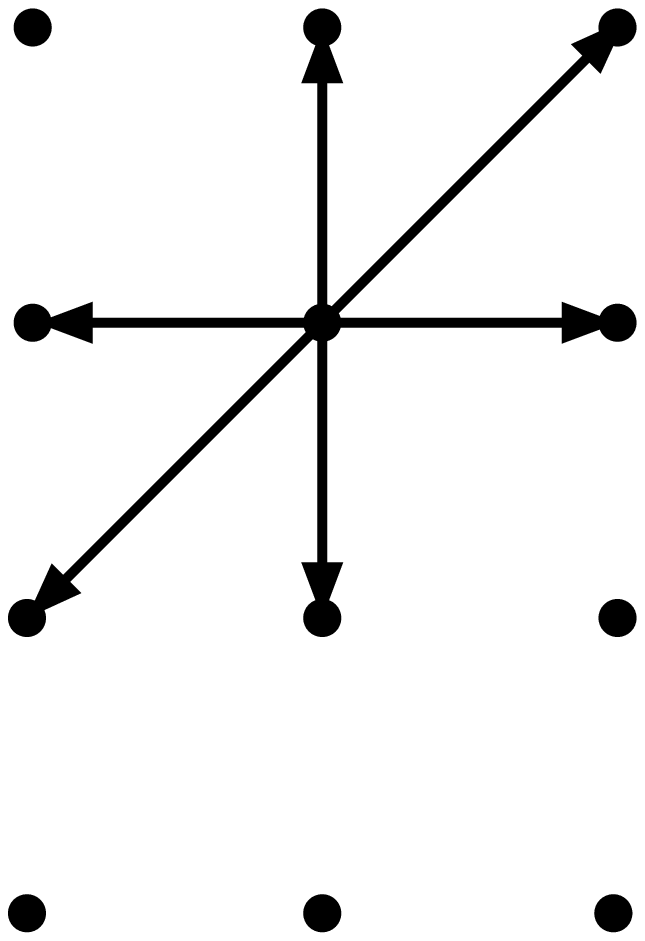} \\
$dP_1=\Hirz[1]$ & $dP_2$ &$dP_3$
\end{tabular}
\caption{\textsl{The toric diagrams of del Pezzo surfaces $dP_n$ with
$n=1,2,3$}} \label{delpezzo}
\end{figure}

There are ten topologically different del Pezzo surfaces. Nine of them are
blow-ups of $\C P^2$ at $n=0,...,8$ points. These surfaces are denoted by
$dP_n$. Additionally, there is $\Hirz[0]=\C P^1 \times \C
P^1$ which is also a Hirzebruch surface. 
Del Pezzo surfaces are completely classified by their Euler characteristic, except for the case $\chi = 4$, where there are the two del Pezzo surfaces $dP_1$ and $\C P^1 \times \C
P^1$ \cite{Hubsch:1992nu}.
Note that for $n \le 3$, $dP_n$ is a toric variety. Their toric
diagrams are given in Figure \ref{delpezzo} and \ref{rootsystemCP2}. For the remainder of this appendix, we focus on the del Pezzo surfaces $dP_n$. The surface $\Hirz[0]=\C P^1 \times \C P^1$ is discussed together with the other Hirzebruch surfaces in Appendix \ref{Hirzebruchsurfaces}.

The hyperplane divisor $H$ of $\C P^2$ has self-intersection $H^2=1$. Each
blow-up introduces one further exceptional divisor $E_i$. Thus, the dimension of the middle homology of a del Pezzo surface $dP_n$ is given by
\begin{equation}
\dim H_2(dP_n,\Z) = n+1\ .
\end{equation}
It is a well-known
fact that exceptional divisors at smooth points of a surface have
self-intersection number $E_i \cdotp E_j = - \delta_{ij}$. Furthermore, a
hypersurface in $\C P^2$ generically does not meet the blown-up points. As a result,
the hyperplane divisor $H$ does not intersect exceptional divisors, $H \cdotp
E_i = 0$. Therefore, we find the following intersection pattern of del Pezzo
surface $dP_n$:
\begin{equation}
H \cdotp H = 1\ , \hspace{2cm} E_i \cdotp H = 0\ , \hspace{2cm} E_i\cdot E_j =
-\delta_{ij}\ ,
\label{dPintersectionpattern}
\end{equation}
with $i=1,...,n$.

We can obtain the canonical divisor $K_{dP_n}$ of a del Pezzo
surface if we take the blown-up cycles into account. It is well-known that the canonical line bundle of a 
manifold $\tilde{Y}$ blown up at a smooth point is given by $[K_{\tilde{Y}}] = \sigma^*[K_Y]\otimes [(\dim Y - 1)E]$ where $\sigma: \tilde{Y} \rightarrow Y$ is the blow-up
and $[E]$ denotes the line bundle corresponding to the exceptional divisor
$E$ \cite{Hubsch:1992nu}.
Iteratively blowing up the exceptional divisors, we find
\begin{equation*}
[K_{dP_n}] = \sigma_n^*[K_{\C P^2}] \otimes \bigotimes_{i=1}^{n} [E_i]\ ,
\end{equation*}
where $\sigma_n: dP_n \rightarrow \C P^2$ is the blow-up at $n$ points.
For the corresponding divisors this reads
\begin{equation*}
K_{dP_n} = -3H + \sum_{i=1}^{n} E_i \ .
\end{equation*}
Here we used $\sigma_n^*[K_{\C P^2}] = [-3H]$.

\subsection{Hirzebruch surfaces} \label{Hirzebruchsurfaces}

\begin{figure}
\begin{tabular}{c@{\hspace{3cm}}c@{\hspace{3cm}}c}
\includegraphics[height=4cm]{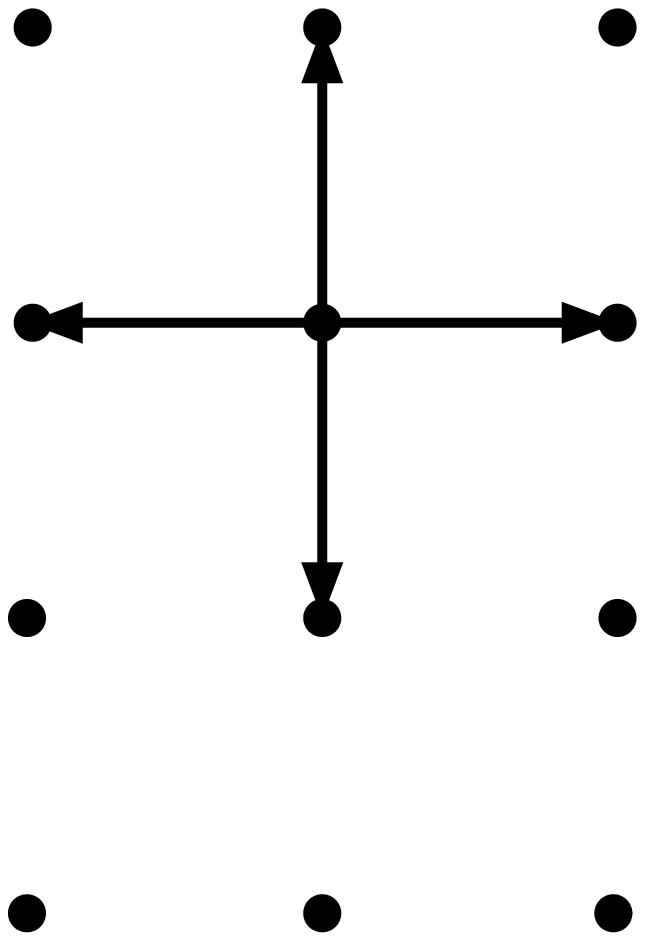}  &
\includegraphics[height=4cm]{F1.eps}  &
\includegraphics[height=4cm]{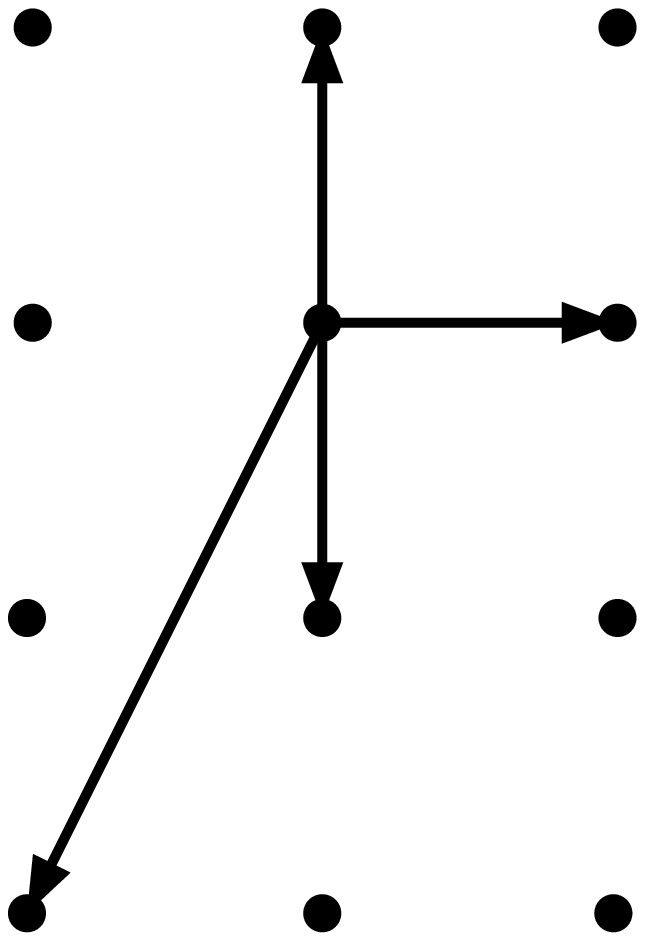} \\
$\Hirz[0]$ & $\Hirz[1]=dP_1$ &$\Hirz[2]$
\end{tabular}
\caption{\textsl{The toric diagrams of Hirzebruch surfaces in the case of
$n=0,1,2$}} \label{hirzebruch}
\end{figure}

Now we turn to Hirzebruch surfaces
$\Hirz[n]$. They are $\C P^1$ fibrations over $\C P^1$. All of them are toric varieties, and we display the fans of the first three Hirzebruch surfaces in Figure \ref{hirzebruch}.
In general, the fan of the $n$-th Hirzebruch surface $\Hirz[n]$ is given
by the cones corresponding to the vectors \cite{Hori:2003ic}
\begin{equation*}
v_1 = \left(\begin{aligned} 1 \\ 0 \end{aligned}\right)\ , \quad v_2 =
\left(\begin{aligned} -1 \\ -n \end{aligned}\right)\ , \quad v_3 =
\left(\begin{aligned} 0 \\ 1 \end{aligned}\right)\ , \quad v_4 = \left(\begin{aligned}
0 \\ -1 \end{aligned}\right) \ ,
\end{equation*}
with $n \in \mathbb{N}$.
Thus the $\C^*$-actions are
\begin{eqnarray*}
\C^*_\lambda: && (z_1,z_2,z_3,z_4) \rightarrow
(\lambda z_1,\lambda z_2,\lambda^n z_3,z_4) \ , \\
\C^*_\mu: && (z_1,z_2,z_3,z_4) \rightarrow (z_1,z_2,\mu z_3,\mu z_4) \ ,
\end{eqnarray*}
and the exceptional set is
\begin{equation}
Z_{\Hirz[n]} = \left\{ (z_1,z_2,z_3,z_4)\in \C^4 | (z_1,z_2) = 0 \text{ or }
(z_3, z_4) = 0\right\} \ .
\end{equation}

Let us now turn to the toric divisors of Hirzebruch
surfaces. The linear equivalences between the toric divisors $T_i$ with
$i=1,...,4$ of $\Hirz[n]$ are given by
\begin{equation}
T_1 = T_2 \quad \textrm{and} \quad T_3=nT_1+T_4 \ .
\end{equation}
and the middle homology group of Hirzebruch surfaces
$H_2(\Hirz[n])$ is generated by two 2-cycles satisfying
\begin{equation}
T_1^2=0\ , \quad T_1T_4 = 1\ , \quad T_4^2=-n \ . \label{HirzebruchDivisors}
\end{equation}
One can calculate the first Chern class
\begin{equation} \label{c1Hirz}
c_1(\Hirz[n])=\sum_{i=1}^4T_i=(2+n)T_1+2T_4 \ ,
\end{equation}
and the Euler characteristic
\begin{equation}
\chi(\Hirz[n]) = \int_{\Hirz[n]} c_2(\Hirz[n]) 
= \sum_{i<j} T_iT_j = 4 \ .
\end{equation}
Note that the first Chern class is positive definite if $n=0,1$. This means
that $\Hirz[0]$ and $\Hirz[1]$ are also del Pezzo surfaces. Indeed, we already
discussed this issue in the previous subsection. In the case of $n=2$, the first Chern class is
positive semi-definite.

Next, we turn to the investigation of curves in Hirzebruch surfaces. Let
$\cal H$ be such a curve of degree $(a,b)$, i.e.\ given by the zero
locus of a polynomial $p$ that is homogeneous of degree $a$ with respect to
$\C^*_\lambda$ and homogeneous of degree $b$ w.r.t.\ $\C^*_\mu$. Then $\cal H$ is
linearly equivalent to $aT_1+bT_4$. Thus we find
\begin{equation}
c_1({\cal H}) = c_1(\Hirz[n])-{\cal H}=(2+n-a)T_1+(2-b)T_4\ .
\end{equation}
Using the intersection numbers \eqref{HirzebruchDivisors}, the Euler
characteristic of $\cal H$ then turns out to be
\begin{equation}
\chi({\cal H}) = \left[2(1-a)-n(1-b)\right]b+2a\ .
\end{equation}
Hence the genus of $\cal H$ is
\begin{equation}
g_{(a,b)}=1-\frac{\chi}{2}=1+b\left[(a-1)+\frac{n}{2}(1-b)\right]-a\ .
\end{equation}
We can apply this result to the case of an O-plane $\cal H$ in $\Hirz[n]$.
In the weak coupling limit of F-Theory on $\Hirz[n]$, it is described
by a curve of degree $(2n+4,4)$ \cite{Morrison:1996na}. Plugging this into
the above equation we obtain $g_{(2n+4,4)}=9$, independent of $n$. We use this result in Table \ref{summExam} in Section~\ref{DonO}.

\section[Classification of non-symplectic involutions of ${\bf K3}$ surfaces]{Classification of non-symplectic involutions of $K3$ surfaces} \label{nikulinClassification}
Here we give a short review on the classification of $K3$ surfaces $X$ equipped with a non-symplectic involution $\iota: X \rightarrow X$, obtained by Nikulin in \cite{Nikulin:1979,Nikulin:1983,Nikulin:1986}.

Such pairs $(X, \iota)$ are classified in terms of a characteristic triplet $(r,a,\delta)$, where $r$ and $a$ are non-negative integers and $\delta$ is zero or one. We explain now how these numbers are defined. For that, we define $S_X$ to be
the Picard lattice of $X$ and $S$ to be the sublattice spanned by
$(1,1)$-forms that are even under the pullback $\iota^*$:
\begin{equation}
S := \left\{ \omega \in H^{(1,1)}(X) | \iota^* \omega = \omega\right\} \ .
\end{equation}
The numbers in the characteristic triplet can be obtained from the structure of
$S$. The first is $r:= \text{rank } S$. Furthermore, for each lattice there exists a dual
lattice $S^*=\text{Hom}(S,\Z)$. Each element $\alpha$ of $S^*$ can be
represented by an element $c$ of $S \otimes \R$ by identifying
$\alpha(\cdot)=(c,\cdot)$, where $(\cdot,\cdot)$ denotes the scalar product on
$S$. Since $(S,S) \subset \Z$ it is obvious that $S \subset S^*$. Indeed, it can
be shown that $S^*/S = (\Z_2)^a$ for an integer $a$ which is the
second entry of the characteristic triplet.
The third entry $\delta$ is defined as follows: Identify every linear form 
$(x, \cdot) \in S^*$ with the corresponding element $x \in S \otimes \R$ and determine $(x,x)$.
If for all such linear forms there is $(x,x) \in \Z$, then $\delta = 0$.
Otherwise, we set $\delta=1$. 
These three numbers $(r,a,\delta)$ determine the pair
$(X,\iota)$ is up to isomorphisms.

There is only a limited set of surfaces $Y$ that arise from $K3$ surfaces $X$
by modding out a non-symplectic involution $\iota: X \rightarrow X$. 
The surface $Y$ is either an Enriques surface (which is the only case with
no fixed point locus), or a rational surface, cf.\ Appendix \ref{complexSurfaces}.
Such manifolds $Y$ naturally admit a $K3$ double cover. 
It was shown by Nikulin that these surfaces correspond to so-called
non-singular DPN-pairs $(Y,C)$. By definition, $Y$ is a non-singular projective
algebraic surface with $b^1(Y)=0$ and $C$ is a non-singular effective divisor
$C=-2K_Y$ in $Y$.\footnote{The definition may be extended to singular effective
divisors $C$. See \cite{alexeev:2004} for a treatment of such cases.}
In this correspondence, $C$ denotes the fixed point locus $X^\iota$ of $\iota$ in $X$. In orientifold models, $C$ is the position of the O-plane.

An important result of Nikulin is that the numbers $(r,a,\delta)$ immediately
give certain properties of the fixed point locus $C\equiv X^\iota$. If the characteristic triplet is not $(10,10,0)$ or $(10,8,0)$,
the O-plane curve $C$ is of the form
\begin{equation}
C=C_g+\sum_{i=1}^k E_i\ ,
\end{equation}
where $C_g$ is a curve of genus $g$ and the $E_i$ are rational curves, that is,
of genus 0. The quantities $g$ and $k$ can be expressed in terms of the
characteristic triplet:
\begin{equation}\label{Nikulin}
\begin{aligned}
g&=\tfrac{1}{2} (22-r-a) \ , \\
k&=\tfrac{1}{2}(r-a)\ . 
\end{aligned}
\end{equation}
It is an important fact that $C_g$ and all $E_i$ do not intersect each other.

An application for the above discussion is given in Appendix \ref{reducible}, where we analyze the case $Y=\Hirz[4]$. It is known \cite{alexeev:2004} that the characteristic
triplet is $(2,0,0)$ in this case. Thus the O-plane is the disjoint union of a
curve of genus 10 and a single rational curve.

\section{The Lie algebra of automorphisms for toric varieties} \label{toricAut}

When talking about possible deformations of a hypersurface $H$ embedded in a
complex manifold $Y$, we always encounter the question which deformations can be
undone by applying an automorphism of the embedding space $Y$. An easy example
can be given by the complex line in $Y=\C P^2$, that is, let $H$ be given by the
zero locus of a homogeneous polynomial of degree one. Fixing the overall
scale factor by setting one coefficient equal to one, $H$ reads
\begin{eqnarray}
H: && z_1 + \alpha z_2 + \beta z_3 = 0\ .
\end{eqnarray}
Naively one might think that the moduli space of $H$ is two-dimensional due to the
two coefficients $\alpha$ and $\beta$. However, this is not the case as the topology of $\C P^2$
stays unchanged if an element of the automorphism group $PGl(3,\C)$ is applied on the
homogeneous coordinates. In other words, we can use automorphisms of the
embedding space to set coefficients to zero, i.e.\ these
coefficients do not represent degrees of freedom. The Lie algebra of
$PGl(3,\C)$ is eight-dimensional. Thus, the moduli space of the
complex line is zero-dimensional and $H$ is unique in $\C P^2$.

It can be difficult to determine the automorphism group for more complicated
manifolds. However, for toric varieties Demazure analyzed the automorphism group in detail and determined the dimension of its algebra \cite{Demazure:1970}.
In this appendix we explain how to determine the dimension of the Lie
algebra of automorphisms, $\dim \aut_T$, in the case of a general
toric variety $T$. 

\begin{figure} \hspace{1.7cm}
\begin{tabular}{c@{\hspace{3cm}}c}
\includegraphics[height=4cm]{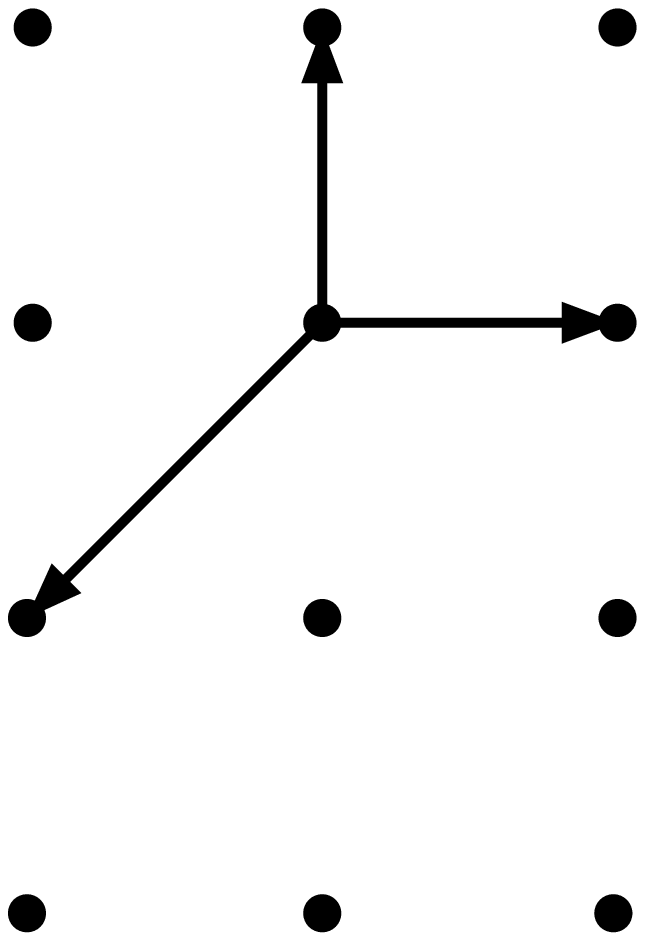}  &
\includegraphics[height=4cm]{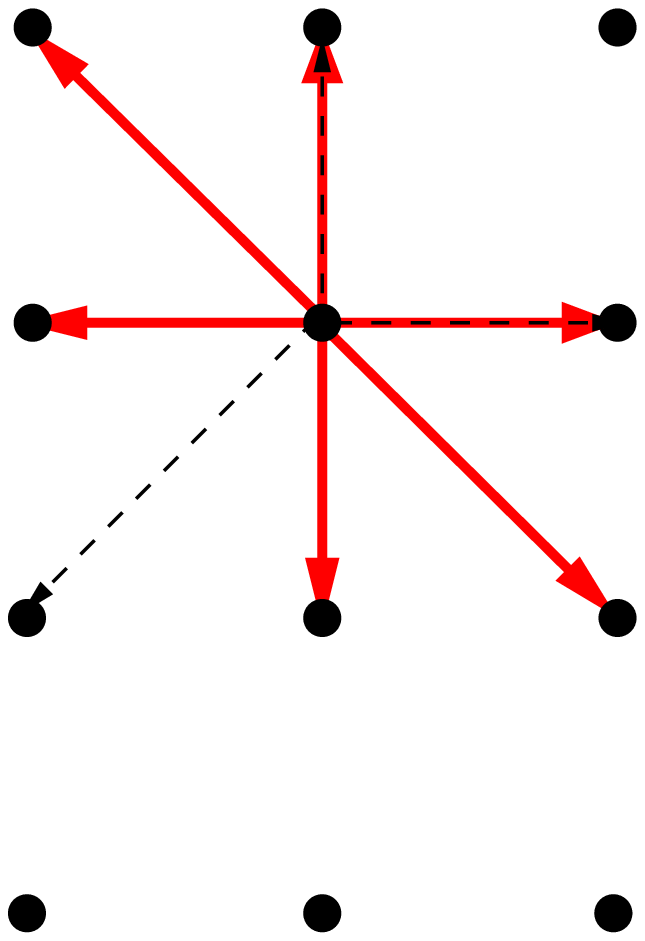} \\
the fan of $\C P^2$ & the same fan with its root system
\end{tabular}
\caption{\textsl{The toric diagram and root system of $\C P^2$}} 
\label{rootsystemCP2}
\end{figure}

Given the fan $\Sigma$ of $T$, we denote the set of $j$-dimensional
cones in $\Sigma$ by $\Sigma(j)$. For any one-dimensional cone $\rho \in
\Sigma(1)$ there is a primitive vector $n(\rho)$ generating $\rho$. A
primitive vector to a cone $\rho \in \Sigma(1)$ is the lattice vector $n(\rho)$
that spans $\rho$ such that there is no other lattice vector $m \neq n(\rho)$
for which $n = a m$ with $a \in \mathbb{N}$. Mapping any cone $\rho \in \Sigma(1)$ to its primitive vector $n(\rho)$ defines an embedding of $\Sigma(1)$ into a
lattice $N$. Similarly, the complete fan $\Sigma$ can be embedded into $N$. We denote the dual lattice of $N$ by $M$ and the natural Cartesian scalar product on $M \otimes N$ by $(\cdot,\cdot)$.
Furthermore, we introduce the root system $R(N,\Sigma)$ of a toric variety.
Abstractly, this is defined as the set of all elements $\alpha$
of the dual lattice $M$ for which exactly one cone $\rho_\alpha \in \Sigma(1)$
with $(\alpha,n(\rho_\alpha))=1$ exists and $(\alpha,n(\rho)) \leq 0$ holds for
all other cones $\rho \in \Sigma(1)$ \cite{Oda:1988}. 
\rem{More formally, this may be rephrased as
\begin{equation}
R(N,\Sigma):=\left\{ \alpha \in M \vert \exists \rho_\alpha \in \Sigma(1):
(\alpha, n(\rho_\alpha))=1 \wedge \forall \rho \in \Sigma(1)\setminus
\{\rho_\alpha\}:(\alpha, n(\rho)) \leq 0 \right\}\ .
\end{equation}
}
This can be nicely illustrated for toric varieties of complex dimension
two. For example, take the fan of $\C P^2$ as given in the left diagram of Figure
\ref{rootsystemCP2}. The corresponding root system is given by the six vectors drawn in the right diagram of Figure \ref{rootsystemCP2}.

Now the theorem due to Demazure
\cite{Demazure:1970} states that for a compact, non-singular toric variety
the dimension of the automorphism algebra is given by
\begin{equation}
\dim \aut_T = \text{rank } N + \# R(N,\Sigma)\ ,
\end{equation}
where $\# R(N,\Sigma)$ is the number of elements of the root system $R(N,\Sigma)$. Applying this theorem to the
case of compact non-singular complex surfaces, we can use that its toric fan
can be embedded into a two-dimensional lattice $N$. Therefore we have $\text{rank
} N = 2$ so that we obtain
\begin{equation} \label{dimautomorphismgroup}
\dim \aut_T = 2 + \# R(N,\Sigma)\ .
\end{equation}
Thus we can deduce the dimension of the automorphism algebra by finding the number of roots. In the case
of $\C P^2$ we obtain six root vectors and hence
\begin{equation}
\dim \aut_{\C P^2} = 2 + 6 = 8\ ,
\end{equation}
which is in agreement with the fact that the automorphism group of $\C P^2$ is $PGl_3(\C)$. However, \eqref{dimautomorphismgroup} holds in general and thus can be applied to any other complex surface. In
particular, for Hirzebruch surfaces it yields
\begin{equation}
\dim \aut_{\Hirz[n]} = \Big\{ \begin{array}{cl} 6 & \text{for $n=0$} \ ,
\\ 5+n & \text{for $n>0$} \ . \end{array}   \label{autF}
\end{equation}
We use these results in Table \ref{summExam} in Section~\ref{DonO} when we count the degrees of freedom for deformations in the F-theory model.

\section[An example: The weak coupling limit for base space
${\Hirz[4]}$]{An example: The weak coupling limit for base space
$\bf \Hirz[4]$} \label{reducible}

We have seen in Section~\ref{moreDoffO} that the suggested counting of F-theory 3-cycles holds in the case
of smooth Calabi-Yau threefolds. In this appendix, we want to discuss the weak
coupling limit of F-theory over a base $\Hirz[4]$. This surface is rational and
has a non-singular effective divisor $C = -2K_{\Hirz[4]}$. Thus it can be
obtained as the quotient space of a $K3$ surface $X$, where the quotient map is
a non-symplectic involution (see Appendix~\ref{nikulinClassification} for a
short review on the theory of such surfaces). However, the elliptically fibred
Calabi-Yau threefold is singular. 

Let us start with the F-theory perspective. We recall the Weierstrass equation 
\begin{equation}
y^2=x^3+fx+g \ ,
\end{equation}
where $f$ and $g$ are sections of $L^{\otimes 4}$ and $L^{\otimes 6}$,
respectively. Here $L$ is a line bundle defined by the Calabi-Yau condition
$c_1(L) = c_1(\Hirz[4])$. As discussed in Appendix~\ref{complexSurfaces}, the
first Chern class is given by
\begin{equation}
c_1(\Hirz[4])=6T_1+2T_4 \ ,
\end{equation}
and therefore we find
\begin{equation}
L=[T_1]^{\otimes 6}\otimes[T_4]^{\otimes 2}\ .
\end{equation}
This means that $f$ and $g$ are polynomials of degree $(24,8)$ and $(30,12)$.
We can write the general form of $f$ and $g$:
\begin{eqnarray}
f &=&
\left(\sum_{\alpha=0}^{6}\sum_{\beta=0}^{24-4\alpha}F_{\alpha\beta}z_1^\beta
z_2^{24-4\alpha-\beta}z_3^\alpha z_4^{6-\alpha}\right)z_4^2 \ , \\
g &=&
\left(\sum_{\alpha=0}^{7}\sum_{\beta=0}^{30-4\alpha}G_{\alpha\beta}z_1^\beta
z_2^{30-4\alpha-\beta}z_3^\alpha z_4^{7-\alpha}\right)z_4.
\end{eqnarray}
It was observed in \cite{Morrison:1996pp} that this elliptically fibred
Calabi-Yau threefold has generically a $D_4$ singularity. By Bertini's Theorem
\cite{Griffiths:1994}, this singularity is located at the base locus of the
threefold, which is $x=y=z_4=0$. Note that the base locus describes a
hyperplane in the base $\Hirz[4]$ given by $z_4=0$. Therefore we expect a
brane at $T_4$ that carries an $SO(8)$ gauge group. With this in mind we now turn
to the weak coupling limit.

In the weak coupling limit, the O-plane $\op$ is described by the zero locus
of a polynomial $h$ of degree $(12,4)$. Its general form is
\begin{equation}
h=\left(\sum_{\alpha=0}^{3}\sum_{\beta=0}^{12-4\alpha}H_{\alpha\beta}z_1^\beta
z_2^{12-4\alpha-\beta}z_3^\alpha z_4^{4-\alpha}\right)z_4 \ .
\end{equation}
Note that $\op$ is reducible. In particular, we find $\op = \op'+ T_4$.
This means that one O-plane splits from $\op$ and coincide with $T_4$. Doing the
same analysis with the D-brane $\cal D$ which is given by the zero-locus of
$\eta^2+12h\chi$ we obtain that $\cal D$ is reducible as well. In particular, we
find ${\cal D} = {\cal D}'+4T_4$. We see that one O-plane and four D-branes
coincide with $T_4$ producing an $SO(8)$ gauge group on $T_4$ in
agreement with the results obtained in the F-theory picture.

By Nikulin's classification \cite{Nikulin:1986} we know that $\Hirz[4]$ has the
characteristic triplet $(2,0,0)$ and by the results in Appendix~\ref{nikulinClassification} this means that the fixed point locus, i.e.\ the
O-plane, is of the form
\begin{equation}
\op = C_{10} + E_1 \ ,
\end{equation}
where $C_{10}$ is a curve of genus 10 and $E_1$ is a rational curve. This fits
with our results above by identifying $\op'=C_{10}$ and $T_4=E_1$. Note
that $\op$ is non-singular meaning that $C_{10}$ and $E_1$ do not intersect.

As we have seen, although the Calabi-Yau threefold is singular, F-theory on
$\Hirz[4]$ has a weak coupling limit. For $B=\Hirz[n]$ with
$n>4$ the corresponding elliptically fibred Calabi-Yau threefold has singularities of $E$-type which cannot appear in an orientifold model in perturbative type IIB . Indeed, $\Hirz[n\geq 5]$ do not have a $K3$ double
cover and thus there cannot exist a dual type IIB orientifold model. The base
$B=\Hirz[3]$ generically has an $A_3$-singularity, which might be obtained
in the perturbative type IIB orientifold model. However, $\Hirz[3]$ apparently does not admit a $K3$
double cover since the $\op$ hypersurface one obtains in the weak
coupling limit is always singular \cite{Nikulin:1986}.

\section{D-brane Deformations and Variations of the Intersection Divisor}
\label{abelSection}

In this appendix, we analyze how the degrees of freedom of D-branes are
distributed between cycles that change the location of intersection points with the
O-plane and those that do not. Our main tool is Abel's Theorem \cite{Griffiths:1994}. We
concentrate on the case where at least one D-brane coincides with the
O-plane. This allows us to ignore additional constraints coming from the
double intersection property of generic allowed D-branes.

The D-brane is described by a section of the line bundle $[-2nK_B]$ in the base space. This bundle
restricts to a bundle $[L]=[-2nK_B]|_O$ on O-plane. The sections of the bundle $[L]$ determine the intersections
between the D-brane and the O-plane. 

Knowing this, we take the following point of view. We treat the O-plane
as a Riemannian surface $O$ of genus $g(O)$ marked with $I$ points. These points are the intersection 
points $P_\alpha$ with the D-brane. They are given by the vanishing locus of a section of $[L]$. In other words, 
the set of points marking $O$ is the divisor $L$ associated to $[L]$. In this picture, varying the intersections points 
corresponds to varying the section in $[L]$. This means that the corresponding divisor $L$ can only change to a linearly 
equivalent divisor $L'$, that is $L'= L + \delta L$, where $\delta L$ is a divisor of a meromorphic function. 
Writing $L=\sum_\alpha P_\alpha$ and $L'=\sum_\alpha P'_\alpha$ yields
\begin{equation}
\delta L = \sum_\alpha (P'_\alpha-P_\alpha).
\end{equation}
By Abel's Theorem we know that
\begin{equation}
{\cal J}(\delta L)=\sum_\alpha \left( \begin{array}{c} \int_{P_\alpha}^{P'_\alpha} \omega_1
\\ \vdots \\ \int_{P_\alpha}^{ P'_\alpha} \omega_g \end{array} \right) \equiv 0
\text{ mod } \Lambda. \label{abelJacobi}
\end{equation}
Here $\cal J$ is the Abel-Jacobi map, $\{\omega_i\mid i=1,...,g\}$ is a basis of
holomorphic 1-forms on $\op$ and $\Lambda$ is the period lattice. Fixing 
the divisor $L$ allows us to consider $\cal J$ as a map $O \rightarrow O$. In following we want to consider infinitesimal variations of
the intersection locus, that is, $P'_\alpha=P_\alpha+\delta P_\alpha$. In this
case, it is sufficient to consider a neighborhood isomorphic to $\C$ around
each point $P_\alpha$ in $L$ and then
\begin{equation}
{\cal J}: \C^I \rightarrow \C^g.
\end{equation}
We now can determine the differential of the Abel-Jacobi map $\cal J$, which
is
\footnote{Here we use the notation of \cite{Griffiths:1994}. Let $\omega_i$ be
a holomorphic 1-form of $\op$. Then write in local coordinates $\omega_i =
f_i(z)dz$. In local coordinates the differential of the path integral is
\begin{equation*}
\frac{d}{dP'_j}\left( \int_{P_j}^{P'_j} \omega_i\right) \Bigg| _{P_j} =
\frac{d}{dP'_j}\left( \int_{P_j}^{P'_j} f_i(z)dz \right) \Bigg|_{P_j}=f(P_j)
\end{equation*}
In this sense, we write $\frac{\omega_i}{dP'_j}\Big|_{P_j} = f(P_j)$.}
\begin{equation}
D{\cal J}(L)=\left( \begin{array}{ccc}
\frac{\omega_1}{dP'_1}\Big|_{P_1} & \cdots & \frac{\omega_1}{dP'_I}\Big|_{P_I}
\\
\vdots & \ddots & \vdots \\
\frac{\omega_g}{dP'_1}\Big|_{P_1} & \cdots & \frac{\omega_g}{dP'_I}\Big|_{P_I}
\end{array}\right)=0. \label{differential}
\end{equation}
This is just the Taylor expansion of eq. (\ref{abelJacobi}) for
infinitesimal variations:
\begin{equation}
{\cal J}(\delta L) = D{\cal J}(L) \left(\begin{array}{c} \delta
p_1 \\ \vdots \\
\delta p_I \end{array} \right)
\end{equation}
By the constraint (\ref{differential}) we find that all allowed variations are
in the kernel of $D{\cal J}(L)$. Thus the number of degrees of
freedom encoded in varying the intersection points is $\dim_\C \ker D{\cal
J}(L)$. By using Riemann-Roch it is not difficult to show that
$D{\cal J}(L)$ has full rank, if $\text{deg } L = I > \text{deg } K_O$. If this is the case, for genus $g(O)>2$, the 
dimension of the kernel is just
\begin{equation}
\dim_\C \ker DJ(L) = I - g(O). \label{movingintersections}
\end{equation}

It is interesting to note the implications of this constraint for special cases. First, let us consider the case of three D-branes coinciding with the O-plane. Then the line bundle $[L]$ is the normal bundle of the O-plane $\nbo$, as expected by geometric intuition. In Eq. (\ref{selfintersection}), we found that $\text{deg} \nbo = 4g(O)-4$ so Eq.(\ref{movingintersections}) is valid. In particular, we find that there are $I-g(O)=3g(O)-4$ degrees of freedom that vary the intersection points. Recall that a single D-brane is a Riemannian surface with $3g(O)-3$ complex deformation degrees of freedom. Thus, fixing all intersections, we are left with just one complex degree of freedom that deforms the D-brane.

Let us illustrate this in the case of $B=\C P^2$. The O-plane as well as the
D-brane is described by a polynomial of degree 6, denoted by $h$ and $d$.
respectively. Suppose that the overall rescaling is fixed in $d$. In this case
we may write
\begin{equation}
d=\alpha h + \tilde{d}_\alpha, \label{specialform}
\end{equation}
where $\alpha \in \C$ and $\tilde{d}$ is chosen such that the equality in eq.
(\ref{specialform}) holds. On the O-plane we have $h=0$ and thus the
intersection divisor is independent of $\alpha$. This is exactly the complex
degree of freedom that deforms the D-brane but leave the intersection divisor
unchanged.

In chapter \ref{moreDoffO} we encountered situations, in which the D-brane locus has $10g(\op)-10$ complex degrees of freedom. This is the case of two recombined
D-branes. Then we find \mbox{$I=(-4K_B)\cdot(-2K_B)=8g(\op)-8$}, so that
\begin{equation}
\dim_\C \ker D{\cal J}(D) = I - g(\op) = 7g(\op)-8.
\end{equation}
This is the number of degrees of freedom that will change the intersection locus
and hence $3g(\op)-2$ degrees of freedom will leave it invariant. Again, this
may be checked nicely in the case of $B=\C P^2$. The recombined D-brane is
determined by a polynomial $d$ of degree 12. Suppose the overall rescaling is
fixed. then we write
\begin{equation}
d = h d_6 + \tilde{d}.
\end{equation}
The polynomial $d_6$ is of degree 6. As before, on the O-plane $h=0$ and thus
$d_6$ has no influence on the intersection locus. It has $7\cdot8/2=28$ complex
coefficients and therefore the same number of complex degrees of freedom. Since
on $\C P^2$ we have $g(\op)=10$, this fits exactly with the number of
$3\cdot 10-2=28$ complex degrees of freedom obtained by Abel's Theorem.

The same considerations apply in the case of $B=\Hirz[0]$. Recall, that in
this case $h$ is a polynomial of degree $(4,4)$ and and $d$ is of degree
$(8,8)$. Additionally, we have $g=9$ yielding by Abel's Theorem $25$ complex
degrees of freedom that leave the intersections invariant. This is precisely the
number of coefficients of a polynomial of degree $(4,4)$.


\begin{thebibliography}{10}

\bibitem{Blumenhagen:2006ci}
R.~Blumenhagen, B.~Kors, D.~Lust and S.~Stieberger, {\it {Four-dimensional
  String Compactifications with D-Branes, Orientifolds and Fluxes}},  {\em
  Phys. Rept.} {\bf 445} (2007) 1--193
  [\href{http://arXiv.org/abs/hep-th/0610327}{{\tt hep-th/0610327}}].

\bibitem{Kachru:2003aw}
S.~Kachru, R.~Kallosh, A.~Linde and S.~P. Trivedi, {\it {De Sitter vacua in
  string theory}},  {\em Phys. Rev.} {\bf D68} (2003) 046005
  [\href{http://arXiv.org/abs/hep-th/0301240}{{\tt hep-th/0301240}}].

\bibitem{Balasubramanian:2005zx}
V.~Balasubramanian, P.~Berglund, J.~P. Conlon and F.~Quevedo, {\it {Systematics
  of moduli stabilisation in Calabi-Yau flux compactifications}},  {\em JHEP}
  {\bf 03} (2005) 007 [\href{http://arXiv.org/abs/hep-th/0502058}{{\tt
  hep-th/0502058}}].

\bibitem{Lust:2005dy}
D.~Lust, S.~Reffert, W.~Schulgin and S.~Stieberger, {\it {Moduli stabilization
  in type IIB orientifolds. I: Orbifold limits}},  {\em Nucl. Phys.} {\bf B766}
  (2007) 68--149 [\href{http://arXiv.org/abs/hep-th/0506090}{{\tt
  hep-th/0506090}}].

\bibitem{Lust:2006zg}
D.~Lust, S.~Reffert, E.~Scheidegger, W.~Schulgin and S.~Stieberger, {\it
  {Moduli stabilization in type IIB orientifolds. II}},  {\em Nucl. Phys.} {\bf
  B766} (2007) 178--231 [\href{http://arXiv.org/abs/hep-th/0609013}{{\tt
  hep-th/0609013}}].

\bibitem{Lust:2005bd}
D.~Lust, P.~Mayr, S.~Reffert and S.~Stieberger, {\it {F-theory flux,
  destabilization of orientifolds and soft terms on D7-branes}},  {\em Nucl.
  Phys.} {\bf B732} (2006) 243--290
  [\href{http://arXiv.org/abs/hep-th/0501139}{{\tt hep-th/0501139}}].

\bibitem{Denef:2005mm}
F.~Denef, M.~R. Douglas, B.~Florea, A.~Grassi and S.~Kachru, {\it {Fixing all
  moduli in a simple F-theory compactification}},  {\em Adv. Theor. Math.
  Phys.} {\bf 9} (2005) 861--929
  [\href{http://arXiv.org/abs/hep-th/0503124}{{\tt hep-th/0503124}}].

\bibitem{Vafa:1996xn}
C.~Vafa, {\it {Evidence for F-Theory}},  {\em Nucl. Phys.} {\bf B469} (1996)
  403--418 [\href{http://arXiv.org/abs/hep-th/9602022}{{\tt hep-th/9602022}}].

\bibitem{Sen:1997bp}
A.~Sen, {\it {Orientifold limit of F-theory vacua}},  {\em Nucl. Phys. Proc.
  Suppl.} {\bf 68} (1998) 92--98
  [\href{http://arXiv.org/abs/hep-th/9709159}{{\tt hep-th/9709159}}].

\bibitem{Beasley:2008dc}
C.~Beasley, J.~J. Heckman and C.~Vafa, {\it {GUTs and Exceptional Branes in
  F-theory - I}},  \href{http://arXiv.org/abs/0802.3391}{{\tt 0802.3391}}.

\bibitem{Beasley:2008kw}
C.~Beasley, J.~J. Heckman and C.~Vafa, {\it {GUTs and Exceptional Branes in
  F-theory - II: Experimental Predictions}},
  \href{http://arXiv.org/abs/0806.0102}{{\tt 0806.0102}}.

\bibitem{Donagi:2008ca}
R.~Donagi and M.~Wijnholt, {\it {Model Building with F-Theory}},
  \href{http://arXiv.org/abs/0802.2969}{{\tt 0802.2969}}.

\bibitem{Donagi:2008kj}
R.~Donagi and M.~Wijnholt, {\it {Breaking GUT Groups in F-Theory}},
  \href{http://arXiv.org/abs/0808.2223}{{\tt 0808.2223}}.

\bibitem{Blumenhagen:2008zz}
R.~Blumenhagen, V.~Braun, T.~W. Grimm and T.~Weigand, {\it {GUTs in Type IIB
  Orientifold Compactifications}},  {\em Nucl. Phys.} {\bf B815} (2009) 1--94
  [\href{http://arXiv.org/abs/0811.2936}{{\tt 0811.2936}}].

\bibitem{Marsano:2009ym}
J.~Marsano, N.~Saulina and S.~Schafer-Nameki, {\it {F-theory Compactifications
  for Supersymmetric GUTs}},  {\em JHEP} {\bf 08} (2009) 030
  [\href{http://arXiv.org/abs/0904.3932}{{\tt 0904.3932}}].

\bibitem{Blumenhagen:2009up}
R.~Blumenhagen, T.~W. Grimm, B.~Jurke and T.~Weigand, {\it {F-theory uplifts
  and GUTs}},  {\em JHEP} {\bf 09} (2009) 053
  [\href{http://arXiv.org/abs/0906.0013}{{\tt 0906.0013}}].

\bibitem{Marsano:2009gv}
J.~Marsano, N.~Saulina and S.~Schafer-Nameki, {\it {Monodromies, Fluxes, and
  Compact Three-Generation F-theory GUTs}},  {\em JHEP} {\bf 08} (2009) 046
  [\href{http://arXiv.org/abs/0906.4672}{{\tt 0906.4672}}].

\bibitem{Blumenhagen:2009yv}
R.~Blumenhagen, T.~W. Grimm, B.~Jurke and T.~Weigand, {\it {Global F-theory
  GUTs}},  \href{http://arXiv.org/abs/0908.1784}{{\tt 0908.1784}}.

\bibitem{Jockers:2008pe}
  H.~Jockers and M.~Soroush,
  {\it Effective superpotentials for compact D5-brane Calabi-Yau geometries},
  Commun.\ Math.\ Phys.\  {\bf 290} (2009) 249
  [arXiv:0808.0761 [hep-th]].


\bibitem{Grimm:2008dq}
  T.~W.~Grimm, T.~W.~Ha, A.~Klemm and D.~Klevers,
  {\it The D5-brane effective action and superpotential in N=1
  compactifications},
  Nucl.\ Phys.\  B {\bf 816} (2009) 139
  [arXiv:0811.2996 [hep-th]].

\bibitem{Alim:2009rf}
  M.~Alim, M.~Hecht, P.~Mayr and A.~Mertens,
  {\it Mirror Symmetry for Toric Branes on Compact Hypersurfaces},
  JHEP {\bf 0909} (2009) 126
  [arXiv:0901.2937 [hep-th]].

\bibitem{Jockers:2009mn}
  H.~Jockers and M.~Soroush,
  {\it Relative periods and open-string integer invariants for a compact
  Calabi-Yau hypersurface},
  Nucl.\ Phys.\  B {\bf 821} (2009) 535
  [arXiv:0904.4674 [hep-th]].

\bibitem{Alim:2009bx}
  M.~Alim, M.~Hecht, H.~Jockers, P.~Mayr, A.~Mertens and M.~Soroush,
  {\it Hints for Off-Shell Mirror Symmetry in type II/F-theory
  Compactifications},
  arXiv:0909.1842 [hep-th].

\bibitem{Grimm:2009ef}
T.~W. Grimm, T.-W. Ha, A.~Klemm and D.~Klevers, {\it {Computing Brane and Flux
  Superpotentials in F-theory Compactifications}},
  \href{http://arXiv.org/abs/0909.2025}{{\tt 0909.2025}}.

\bibitem{Aganagic:2009jq}
M.~Aganagic and C.~Beem, {\it {The Geometry of D-Brane Superpotentials}},
  \href{http://arXiv.org/abs/0909.2245}{{\tt 0909.2245}}.


\bibitem{Grimm:2009sy}
  T.~W.~Grimm, T.~W.~Ha, A.~Klemm and D.~Klevers,
  {\it Five-Brane Superpotentials and Heterotic/F-theory Duality},
  arXiv:0912.3250 [Unknown].

\bibitem{Jockers:2009ti}
  H.~Jockers, P.~Mayr and J.~Walcher,
  {\it On N=1 4d Effective Couplings for F-theory and Heterotic Vacua},
  arXiv:0912.3265 [Unknown].


\bibitem{Walcher:2006rs}
  J.~Walcher,
  {\it Opening mirror symmetry on the quintic},
  Commun.\ Math.\ Phys.\  {\bf 276} (2007) 671
  [arXiv:hep-th/0605162].


\bibitem{Morrison:2007bm}
  D.~R.~Morrison and J.~Walcher,
  {\it D-branes and Normal Functions},
  arXiv:0709.4028 [hep-th].


\bibitem{Walcher:2009uj}
  J.~Walcher,
  {\it Calculations for Mirror Symmetry with D-branes},
  JHEP {\bf 0909} (2009) 129
  [arXiv:0904.4905 [hep-th]].

\bibitem{Li:2009dz}
S.~Li, B.~H. Lian and S.-T. Yau, {\it Picard-Fuchs Equations for Relative
  Periods and Abel- Jacobi Map for Calabi-Yau Hypersurfaces},
  \href{http://arXiv.org/abs/0910.4215}{{\tt 0910.4215}}.


\bibitem{Braun:2008ua}
A.~P. Braun, A.~Hebecker and H.~Triendl, {\it {D7-Brane Motion from M-Theory
  Cycles and Obstructions in the Weak Coupling Limit}},  {\em Nucl. Phys.} {\bf
  B800} (2008) 298--329 [\href{http://arXiv.org/abs/0801.2163}{{\tt
  0801.2163}}].

\bibitem{Nikulin:1979}
V.~V. Nikulin, {\it {On factor groups of the automorphism group of hyperbolic
  forms modulo subgroups generated by 2-reï¬ections}},  {\em Soviet. Math.
  Dokl.} {\bf 20} (1979) 1156â1158.

\bibitem{Nikulin:1983}
V.~V. Nikulin, {\it {Quotient-groups of groups of automorphisms of hyperbolic
  forms by subgroups generated by 2-reï¬ections. Algebro-geometric
  applications}},  {\em J. Soviet Math.} {\bf 22} (1983) 1401â1476.

\bibitem{Nikulin:1986}
V.~V. Nikulin, {\it {Discrete reï¬ection groups in Lobachevsky spaces and
  algebraic surfaces}}, . Proceedings of the International Congress of
  Mathematicians, Vol. 1, 2 (Berkeley, Calif., 1986) (Providence, RI).

\bibitem{Morrison:1996na}
D.~R. Morrison and C.~Vafa, {\it {Compactifications of F-Theory on Calabi--Yau
  Threefolds -- I}},  {\em Nucl. Phys.} {\bf B473} (1996) 74--92
  [\href{http://arXiv.org/abs/hep-th/9602114}{{\tt hep-th/9602114}}].

\bibitem{Morrison:1996pp}
D.~R. Morrison and C.~Vafa, {\it {Compactifications of F-Theory on Calabi--Yau
  Threefolds -- II}},  {\em Nucl. Phys.} {\bf B476} (1996) 437--469
  [\href{http://arXiv.org/abs/hep-th/9603161}{{\tt hep-th/9603161}}].

\bibitem{Sen:1996vd}
A.~Sen, {\it {F-theory and Orientifolds}},  {\em Nucl. Phys.} {\bf B475} (1996)
  562--578 [\href{http://arXiv.org/abs/hep-th/9605150}{{\tt hep-th/9605150}}].

\bibitem{Collinucci:2008pf}
A.~Collinucci, F.~Denef and M.~Esole, {\it {D-brane Deconstructions in IIB
  Orientifolds}},  \href{http://arXiv.org/abs/0805.1573}{{\tt 0805.1573}}.

\bibitem{Aspinwall:1996mn}
P.~S. Aspinwall, {\it {K3 surfaces and string duality}},
  \href{http://arXiv.org/abs/hep-th/9611137}{{\tt hep-th/9611137}}.

\bibitem{Collinucci:2008zs}
A.~Collinucci, {\it {New F-theory lifts}},  {\em JHEP} {\bf 08} (2009) 076
  [\href{http://arXiv.org/abs/0812.0175}{{\tt 0812.0175}}].

\bibitem{Collinucci:2009uh}
A.~Collinucci, {\it {New F-theory lifts II: Permutation orientifolds and
  enhanced singularities}},  \href{http://arXiv.org/abs/0906.0003}{{\tt
  0906.0003}}.

\bibitem{Brunner:2003zm}
I.~Brunner and K.~Hori, {\it {Orientifolds and mirror symmetry}},  {\em JHEP}
  {\bf 11} (2004) 005 [\href{http://arXiv.org/abs/hep-th/0303135}{{\tt
  hep-th/0303135}}].

\bibitem{Griffiths:1994}
P.~A. Griffiths and J.~Harris, {\em Principles of algebraic geometry}.
\newblock Wiley, New York [u.a.], 1994.

\bibitem{Candelas:1990pi}
P.~Candelas and X.~de~la Ossa, {\it {Moduli Space of Calabi-Yau Manifolds}},
  {\em Nucl. Phys.} {\bf B355} (1991) 455--481.

\bibitem{Friedman-Morgan(1994)}
R.~Friedman and J.~W. Morgan, {\em Smooth four-manifolds and complex surfaces},
  vol.~27 of {\em Ergebnisse der Mathematik und ihrer Grenzgebiete}.
\newblock Springer, 1994.

\bibitem{Braun:2009wh}
A.~P. Braun, R.~Ebert, A.~Hebecker and R.~Valandro, {\it {Weierstrass meets
  Enriques}},  \href{http://arXiv.org/abs/0907.2691}{{\tt 0907.2691}}.

\bibitem{Green:1987rw}
P.~Green and T.~Hubsch, {\it {Polynomial Deformations and Cohomology of
  Calabi-Yau Manifolds}},  {\em Commun. Math. Phys.} {\bf 113} (1987) 505.

\bibitem{Safarevic}
I.~R. Safarevic, {\em Basic algebraic geometry}.
\newblock Springer, Berlin [u.a.], 1977.

\bibitem{Jockers:2004yj}
H.~Jockers and J.~Louis, {\it {The effective action of D7-branes in N = 1
  Calabi-Yau orientifolds}},  {\em Nucl. Phys.} {\bf B705} (2005) 167--211
  [\href{http://arXiv.org/abs/hep-th/0409098}{{\tt hep-th/0409098}}].

\bibitem{Hori:2003ic}
K.~Hori {\em et.~al.}, {\it {Mirror symmetry}}, . Providence, USA: AMS (2003)
  929 p.

\bibitem{Cordova:2009fg}
C.~Cordova, {\it {Decoupling Gravity in F-Theory}},
  \href{http://arXiv.org/abs/0910.2955}{{\tt 0910.2955}}.

\bibitem{Braun:2008pz}
A.~P. Braun, A.~Hebecker, C.~Ludeling and R.~Valandro, {\it {Fixing D7 Brane
  Positions by F-Theory Fluxes}},  {\em Nucl. Phys.} {\bf B815} (2009) 256--287
  [\href{http://arXiv.org/abs/0811.2416}{{\tt 0811.2416}}].

\bibitem{Mori}
S.~Mori and S.~Mukai, {\it {Classification of Fano 3-folds with $B_2 2$}}, .

\bibitem{Hubsch:1992nu}
T.~Hubsch, {\it {Calabi-Yau manifolds: A Bestiary for physicists}}, .
  Singapore, Singapore: World Scientific (1992) 362 p.

\bibitem{alexeev:2004}
V.~Alexeev and V.~V. Nikulin, {\it Classification of log del pezzo surfaces of
  index $ \le 2 $},  2004.

\bibitem{Demazure:1970}
M.~Demazure, {\it {Sous-groupes algebriques de rang maximum du groupe de
  Cremona}},  {\em Ann. Sci. Ecole Norm. Sup.} {\bf 3} (1970) 507--588.

\bibitem{Oda:1988}
T.~Oda, {\em Convex bodies and algebraic geometry}.
\newblock 442 Ergebnisse der Mathematik und ihrer Grenzgebiete ; Folge 3, Bd.
  15. Springer-Verl., 1988.

\end{thebibliography}
\bibliographystyle{JHEP-2}

\providecommand{\href}[2]{#2}\begingroup\raggedright\endgroup

\end{document}